\pdfoutput=1
\documentclass[letterpaper,titlepage,11pt]{article}

\usepackage{jheppub} 
\usepackage[T1]{fontenc}
\usepackage[utf8]{inputenc}
\usepackage[turkish, english]{babel} 

\usepackage[active]{srcltx}
\usepackage{textcomp}
\usepackage{amsmath}
\usepackage{amssymb}
\usepackage{esint}
\usepackage{multirow}
\usepackage{empheq}
\usepackage{textcomp}
\usepackage{xcolor}

\def\cA{\mathcal{A}}

\def\cF{\mathcal{F}}
\def\cG{\mathcal{G}}

\def\cL{\mathcal{L}}
\def\cM{\mathcal{M}}
\def\cN{\mathcal{N}}

\def\cX{\mathcal{X}}
\def\cY{\mathcal{Y}}

\usepackage{enumerate}
\usepackage{xcolor}
\def\be{\begin{eqnarray}}
\def\ee{\end{eqnarray}}
\def\beann{\begin{eqnarray*}}
\def\eeann{\end{eqnarray*}}
\def\beq{\begin{equation}}
\def\eeq{\end{equation}}
\def\ba{\begin{array}}
\def\ea{\end{array}}
\def\ben{\begin{enumerate}}
\def\een{\end{enumerate}}
\def\bea{\begin{eqnarray}}
\def\eea{\end{eqnarray}}

\providecommand{\Lt}{{\tt L}}
\renewcommand{\Lt}{{\tt L}}

\providecommand{\Mt}{{\tt M}}
\renewcommand{\Mt}{{\tt M}}

\providecommand{\Gt}{{\tt G}}
\renewcommand{\Gt}{{\tt G}}

\providecommand{\St}{{\tt S}}
\renewcommand{\St}{{\tt S}}

\newcommand{\sbk}[1]{\left[ #1\right]}
\newcommand{\cbk}[1]{\left\{#1\right\}}

\def\cA{{\cal A}}

\def\cF{{\cal F}}
\def\cG{{\cal G}}

\def\cL{{\cal L}}
\def\cM{{\cal M}}
\def\cN{{\cal N}}

\def\cX{{\cal X}}
\def\cY{{\cal Y}}

\def\be{\begin{equation}}
\def\ee{\end{equation}}
\def\bea{\begin{eqnarray}}
\def\eea{\end{eqnarray}}
\def\ba{\begin{array}}
\def\ea{\end{array}}

\usepackage{amsfonts}

\usepackage{tikz}

\usepackage{bbm}

\DeclareMathOperator{\extdm}{d}
\newcommand{\extd}{\extdm \!}

\title{\boldmath\LARGE{{$\mathcal{N}=1$ Jackiw --Teitelboim supergravity beyond the Schwarzian regime}}}
\subheader{\today}
\newcommand{\itu}{\dagger}

\author[\itu]{H. T. \"Ozer}
\emailAdd{ozert@itu.edu.tr}
\author[\itu]{,\,\,\,Ayt\"ul Filiz}
\emailAdd{aytulfiliz@itu.edu.tr}

\affiliation[\itu]{Istanbul Technical University,\,Faculty of Science and Letters,
\,Physics Department,\\34469 Maslak,\,Istanbul,Turkey.}

\abstract{
We investigate the asymptotic symmetry structure of two--dimensional dilaton gravity in its $\mathcal{N}=1$ supersymmetric extension based on the $\mathfrak{osp}(1|2)$ Lie superalgebra. Within the BF theoretical framework, we analyze affine and superconformal boundary conditions and systematically derive the corresponding asymptotic symmetry algebra(ASA). While the bosonic theory reproduces the Virasoro algebra and its affine enhancement, the supersymmetric extension yields a classical $\mathcal{N}=1$ superconformal algebra whose realization is dynamically restricted by the dilaton supermultiplet.
We show that the boundary behavior of the dilaton induces a controlled dynamical reduction of the full affine $\mathfrak{osp}(1|2)_k$ symmetry to its $\tt{O}\tt{S}p(1|2)$ stabilizer subalgebra, while simultaneously generating an abelian ideal composed of mutually commuting modes. This establishes a coherent interplay between asymptotic symmetry breaking and symmetry extension in low--dimensional supergravity.
Our construction generalizes previous analyses of $\mathfrak{sl}(2,\mathbb{R})$ dilaton gravity to the supersymmetric setting and provides a consistent bulk--based framework for investigating boundary dynamics beyond the Schwarzian regime.
}

\begin{document}

\maketitle
\flushbottom
\newpage
\section{Introduction}
\label{sec:intro}\vspace{0.5cm}
The holographic principle~\cite{Maldacena:1997re} asserts that a gravitational theory defined in $d$ dimensions can be reformulated through an equivalent boundary theory living in $(d{-}1)$ dimensions. This paradigm inspired investigations into two--dimensional $\mathrm{AdS}_2$ holography in the late 1990s, primarily driven by the analysis of asymptotic symmetries in Jackiw--Teitelboim (JT) gravity~\cite{Teitelboim:1983ux,Jackiw:1984je} and dilaton models arising from dimensional reduction~\cite{Strominger:1998yg,Cadoni:1998sg,Hotta:1998iq,Navarro-Salas:1999zer}. These studies have recently experienced a resurgence, motivated by the conjectured duality between JT gravity and the Sachdev--Ye--Kitaev (SYK) model~\cite{Maldacena:2016hyu,Jensen:2016pah,Sachdev:1992fk,Kitaev}, which exemplifies a solvable system with quantum chaotic behavior. Within this correspondence, the Schwarzian boundary action emerges as a low-energy effective description~\cite{Mertens:2018fds}.

Conformal field theory has emerged as a foundational element in contemporary theoretical physics, with far--reaching applications from string theory to the study of complex many--body systems. Its significance lies in the stringent symmetry principles it enforces, which serve to tightly constrain the landscape of admissible physical models. In two dimensions, conformal invariance proves exceptionally potent, manifesting in the form of the infinite--dimensional Virasoro algebra~\cite{Virasoro:1969zu}. Over time, extensive research has been directed toward extending this symmetry to more generalized algebraic settings, such as the Kac--Moody algebras~\cite{Kac_1968,Moody:1968zz} and $W$--algebras~\cite{Zamolodchikov:1985wn}. Simultaneously, substantial effort has been invested in deepening the theoretical understanding of the Virasoro structure itself~\cite{DiFrancesco:1997nk,Blumenhagen:2009zz,Prochazka:2014gqa,Ozer:2015dha}. These algebraic symmetries also underpin the formulation of two--dimensional dilaton gravity, wherein the appearance of the Virasoro algebra at the boundary offers a natural bridge between conformal techniques and gravitational dynamics.

Two--dimensional pure gravity is identified as a topological theory, owing to the lack of propagating local degrees of freedom. In $\mathrm{AdS}_2$ geometries, the symmetry group is given by $\St\Lt(2,\mathbb{R})$, which functions both as the isometry and asymptotic symmetry group and implies a vanishing central charge~\cite{Almheiri:2014cka}. The incorporation of a dilaton field, however, renders the theory dynamical: the dilaton couples to the metric and activates additional boundary excitations, leading to frameworks such as Jackiw--Teitelboim  gravity~\cite{Teitelboim:1983ux,Jackiw:1984je,Maldacena:2016upp,Jensen:2016pah}. This modification extends the symmetry algebra from $\St\Lt(2,\mathbb{R})$ to the infinite--dimensional Virasoro algebra~\cite{Cadoni:1999ja}, thereby allowing for a nonvanishing central charge. JT gravity, most notably through its Schwarzian boundary term, has become a powerful tool for exploring aspects of quantum gravity, particularly in light of its holographic duality with the SYK model~\cite{Maldacena:2016upp,Engelsoy:2016xyb,Almheiri:2014cka}. These developments have placed two--dimensional dilaton gravity at the core of contemporary investigations into holography, black hole thermodynamics, and quantum information~\cite{Cvetic:2016eiv,Sachdev:2019bjn,Sachdev:2010um}.

The intricate internal structure of JT gravity is fundamentally grounded in its boundary symmetry content and the formal mathematical scaffolding that supports it. The theory maintains a profound connection to conformal symmetry through its interplay with the Virasoro algebra and Schwarzian derivatives~\cite{Iliesiu,Mertens,OvisenkoSchwarzian}, and it has been systematically broadened by the implementation of relaxed boundary conditions and higher--spin extensions~\cite{Gross:2016kjj,Witten:2016iux,Yoon}. The boundary behavior can also be effectively captured using coadjoint orbit methods, which serve as powerful tools for probing symmetry--breaking phenomena and their physical ramifications~\cite{Davison:2016ngz,NarayanYoon,Fu:2016vas,Grumiller:2017qao,Alkalaev:2013fsa,Grumiller:2013swa,Grumiller,Kine,Fukuyama}. While the Schwarzian formalism remains a crucial link to the SYK dual, the scope of asymptotic symmetry analysis extends beyond it; in particular, the BF formulation of JT gravity offers a structurally independent yet equally robust avenue for such inquiries~\cite{Grumiller:2002nm,Banks:1990mk}. Within this formalism, the dilaton field functions as a Lagrange multiplier that enforces curvature constraints, and the associated ASA can be extended, for instance, to a Virasoro--Kac--Moody structure, via the inclusion of internal symmetry sectors~\cite{Ikeda:1993fh,Jackiw:1982hg}. This flexibility underscores the utility of JT gravity as a versatile platform for studying gravitational dynamics in lower dimensions, irrespective of adherence to the Schwarzian perspective. 
This methodological structure aligns with the strategy developed in our previous work on bosonic $\mathfrak{sl}(3,\mathbb{R})$ JT gravity~\cite{Ozer:2025bpb}, and here we implement its supersymmetric analogue using the $\mathfrak{osp}(1|2)$ superalgebra.

While prior analyses of JT supergravity, particularly those centered around the Schwarzian action \cite{Almheiri:2014cka,Maldacena:2016upp,Cadoni:1999ja,Engelsoy:2016xyb,Cvetic:2016eiv,Sachdev:2019bjn,Sachdev:2010um}, focus on the boundary effective theory and its relation to the SYK model, our approach emphasizes the bulk gauge-theoretic formulation and its implications for asymptotic symmetry breaking. In contrast to the boundary-centered derivations of the $\cN=1$ super--Schwarzian theory, we construct the full $\mathfrak{osp}(1|2)$ asymptotic algebra via BF theory and coadjoint orbit methods, thereby uncovering dynamical constraints absent in Schwarzian-only treatments. This strategy is conceptually aligned with our earlier $\mathfrak{sl}(3,\mathbb{R})$ construction \cite{Ozer:2025bpb}, and extends the findings of \cite{Grumiller:2002nm,Banks:1990mk,Ikeda:1993fh,Jackiw:1982hg} into the supersymmetric domain. As such, the present work provides a structurally distinct extension of JT supergravity beyond existing formulations.

In the framework of two--dimensional JT gravity, the Schwarzian formalism is frequently utilized to characterize boundary dynamics and to establish its correspondence with the SYK model~\cite{Grumiller:2002nm,Ikeda:1993fh,Jackiw:1982hg}. Nevertheless, when the objective shifts toward understanding the structure of asymptotic symmetries rather than boundary--specific features, reliance on the Schwarzian approach becomes nonessential~\cite{Grumiller:2002nm,Jackiw:1982hg}. In analogy with the Chern--Simons treatment of three--dimensional gravity and the associated Brown--Henneaux boundary conditions, JT gravity admits a BF--theoretic formulation where the dilaton functions as a Lagrange multiplier enforcing curvature constraints~\cite{Grumiller:2002nm,Banks:1990mk}. This perspective enables the realization of extended ASA's, such as the Virasoro or Virasoro--Kac--Moody structures, via the inclusion of internal symmetries and the imposition of compatible boundary data~\cite{Ikeda:1993fh,Jackiw:1982hg}. In particular, coadjoint orbit techniques~\cite{Ikeda:1993fh,Jackiw:1982hg} provide a systematic route for exploring such algebras, especially in scenarios featuring boundary--induced symmetry breaking. As a result, the BF formulation accentuates the adaptability of JT gravity in addressing holographic and asymptotic structures beyond the confines of the Schwarzian perspective.

The structural sophistication of JT gravity stems from the intricate interaction between its boundary symmetry content and formal mathematical architecture. The model exhibits a deep connection to conformal symmetry through the Virasoro algebra and Schwarzian derivatives~\cite{Iliesiu,Mertens,OvisenkoSchwarzian}, and its scope has been broadened by incorporating relaxed boundary conditions and higher--spin enhancements~\cite{Gross:2016kjj,Witten:2016iux,Yoon}. Within this setting, coadjoint orbit techniques provide a powerful framework for probing symmetry breaking phenomena and their physical ramifications~\cite{Davison:2016ngz,NarayanYoon,Fu:2016vas,Grumiller:2017qao,Alkalaev:2013fsa,Grumiller:2013swa,Grumiller,Kine,Fukuyama}. Although the Schwarzian approach plays a central role in the SYK duality, the study of asymptotic symmetries does not intrinsically depend on it~\cite{Grumiller:2002nm,Banks:1990mk}. The BF formulation is particularly well-suited in this regard, as it seamlessly integrates both affine and conformal symmetry realizations~\cite{Ikeda:1993fh,Jackiw:1982hg}. In the present study, we expand this line of investigation into the realm of $\mathfrak{osp}(1|2)$ supergravity, aiming to construct supersymmetric asymptotic structures that extend beyond the Schwarzian regime. 

This phrasing, however, is not intended to suggest a replacement or extension of the Schwarzian theory at the level of physical content or dynamical observables. Instead, our approach focuses on a methodological generalization: we continue the asymptotic symmetry analysis in the BF formalism, now within the \(\mathfrak{osp}(1|2)\) supersymmetric setting. This construction is conceptually aligned with our earlier $\mathfrak{sl}(3,\mathbb{R})$ study \cite{Ozer:2025bpb}, wherein the coadjoint orbit framework and bulk gauge formulation were used to realize symmetry-breaking mechanisms. The present work provides a supersymmetric parallel to that framework and should be seen as a structural, rather than physical, extension beyond the Schwarzian perspective.

Within AdS$_2$ gravity, boundary conditions are generally classified into two dominant types: the broader affine class and the more constrained conformal class. Each class is associated with a distinct ASA and yields unique boundary dynamics. The affine category naturally arises in the Poisson sigma model, where both the dilaton and the boundary metric are allowed to fluctuate at leading order. This framework gives rise to an infinite--dimensional centerless affine $\mathfrak{sl}(2,\mathbb{R})$ current algebra. While the Schwarzian action can be seen as a projection embedded within this more general construction, the complete theory supports a considerably richer algebraic structure. On the other hand, conformal boundary conditions impose geometric constraints that lead to a low--energy effective Schwarzian theory, describing pseudo--Goldstone modes resulting from spontaneous breaking of reparametrization symmetry. These configurations exhibit an off--shell Virasoro symmetry that reduces to $\St\Lt(2,\mathbb{R})$ on-shell, and are widely regarded as the gravitational duals to the infrared limit of the SYK model. Altogether, the affine and conformal classes of boundary conditions define complementary sectors in AdS$_2$ holography and highlight the conceptual breadth of two--dimensional dilaton gravity beyond the confines of the JT framework.

Unlike the well--established $\mathfrak{sl}(2,\mathbb{R})$ formulation of JT gravity, where boundary symmetries are governed by the linear Virasoro algebra, the $\mathfrak{sl}(3,\mathbb{R})$ extension offers a fertile ground for exploring more complex, nonlinear structures such as the $\mathcal{W}_3$ algebra~\cite{Ozer:2025bpb}. This framework is motivated by the pursuit of higher-spin generalizations of dilaton gravity and by the goal of understanding their impact on boundary dynamics. In particular, the inclusion of spin--3 fields alters the asymptotic symmetry structure: it not only enhances the existing symmetry but also introduces a tunable mechanism of symmetry breaking. This formulation yields fresh perspectives on how two--dimensional gravity responds to generalized boundary conditions and contributes to the broader effort to connect higher--spin JT gravity to SYK--like holographic duals. Furthermore, given that $\mathcal{W}_3$ algebras naturally arise in large--$N$ matrix models and in certain deformed SYK constructions, our setup may provide a plausible dual description for such deformed CFT$_1$ theories.

Recent developments have employed JT gravity models augmented with $\St\Lt(3,\mathbb{R})$ symmetry to explore the influence of nonlinear structures, such as the $\mathcal{W}_3$ algebra, on boundary dynamics~\cite{Ozer:2025bpb}. These investigations have clarified how higher-spin fields deform the underlying ASA and support symmetry-breaking mechanisms. In this work, we extend the analysis beyond the purely bosonic realm by introducing a supersymmetric generalization grounded in $\mathfrak{osp}(1|2)$ supergravity. In this framework, the incorporation of a dilaton supermultiplet leads to a novel family of asymptotic field configurations. This setup not only reconfigures the algebraic structure but also furnishes a natural platform for studying supersymmetry breaking patterns and their implications for boundary dynamics.

While the $\mathfrak{sl}(2,\mathbb{R})$ JT gravity model induces a truncation of the infinite--dimensional Virasoro algebra to its $\St\Lt(2,\mathbb{R})$ subgroup due to the boundary dynamics of the dilaton, the extended framework presented here carries out an analogous symmetry reduction for the $\mathcal{N}=1$ superconformal algebra. In this supersymmetric context, the dilaton and higher-spin fields jointly serve as dynamical stabilizers that restrict admissible asymptotic gauge variations, thereby enforcing a reduction of the full $\mathcal{N}=1$ structure to its $\mathfrak{osp}(1|2)$ subalgebra. This symmetry breaking not only produces a consistent algebraic truncation but also provides a coherent method for analyzing such reductions within higher--spin holography through the BF theory formalism. Throughout this analysis, the $\mathfrak{osp}(1|2)$ subalgebra is understood as the extended realization induced from the affine $\mathfrak{osp}(1|2)_k$ or $\mathcal{N}=1$ $\mathfrak{osp}(1|2)$ superconformal algebra, shaped by the dynamical influence of the dilaton sector.

This work centers on the extension of the ASA arising from symmetry--breaking processes, with a particular focus on the $\mathfrak{osp}(1|2)$ generalization of the JT gravity framework. This setting enables a detailed exploration of how the algebraic structure evolves and how these modifications influence the underlying physical dynamics. Previous studies have demonstrated that while a Virasoro algebra can be reconstructed within two--dimensional dilaton gravity~\cite{Grumiller:2013swa}, the associated conserved charges were not entirely integrable, posing challenges for the formulation of a consistent AdS$_2$/CFT$_1$ correspondence. This issue was subsequently resolved by introducing dilaton--dependent charges and relaxing the assumption of Casimir invariance~\cite{Grumiller:2017qao}, which allowed for the full manifestation of the Virasoro symmetry and reinforced the conformal field theory interpretation. These advancements highlight the critical function of extended boundary conditions in AdS$_2$ gravity and pave the way for deeper inquiries into its holographic structure.

In conclusion, this study undertakes a comprehensive investigation into the relationship between two--dimensional dilaton gravity, boundary conditions, and symmetry structures, aiming to fill key gaps identified in the existing body of research. The structural richness and symmetry features of JT gravity offer new angles from which to approach our understanding of holographic dualities. Within this broader landscape, two--dimensional gravity models have established themselves as foundational elements in theoretical physics, and the insights developed here are anticipated to inform and inspire subsequent explorations.

This paper investigates the asymptotic symmetry structure emerging in a two--dimensional JT supergravity model formulated on the foundation of the Lie superalgebra $\mathfrak{osp}(1|2)$. In this construction, the dilaton field is promoted from a classical scalar to a supersymmetry--sensitive supermultiplet, allowing not only reparametrization symmetries but also their supersymmetric counterparts to act nontrivially at the boundary. Our methodology adopts a BF theory formulation, wherein boundary conditions are leveraged to construct and systematically regulate the breaking of these extended supersymmetries. Anchored in coadjoint orbit techniques, this framework provides a firm basis for realizing supersymmetric holographic dualities that transcend the traditional Schwarzian setting.

The interplay between two--dimensional dilaton gravity and SYK--like systems has recently attracted renewed scholarly focus. In particular, the study presented in~\cite{Momeni:2024ygv} introduces a deformed SYK model as a plausible holographic dual to generalized JT gravity, contending that altered boundary terms and higher-spin extensions may serve to encode new structural elements of the corresponding dual theory.

In a complementary vein, multiple studies, such as~\cite{Momeni:2020tyt,Momeni:2020zkx,Momeni:2021jhk}, have investigated higher--spin extensions of AdS$_2$ gravity, highlighting the central significance of extended symmetry algebras, especially $\mathcal{W}$--algebras, in organizing asymptotic dynamics. The present framework resonates with these contributions by formulating an $\mathfrak{osp}(1|2)$ extension of JT gravity within a BF--theoretic scheme that accommodates both symmetry breaking mechanisms and coadjoint orbit methodology. As a result, our construction yields a tangible realization of generalized boundary dynamics that naturally integrates into the broader context of higher--spin gravitational theories.

In this regard, the present construction is best interpreted as a supersymmetric extension of our previous bulk--based asymptotic symmetry analysis, rather than as a novel dynamical theory beyond the Schwarzian paradigm.

Our construction differs conceptually and technically from conventional JT gravity formulations based on \(\mathfrak{sl}(2,\mathbb{R})\) \cite{Momeni:2024ygv,Momeni:2020tyt,Momeni:2020zkx}, particularly those that emphasize boundary actions, dilaton constraints, or effective Schwarzian descriptions. While those models often encode symmetry breaking through boundary terms and thermodynamic consistency conditions, our approach derives asymptotic symmetries directly from the bulk via the structure of the gauge field and its residual supersymmetries. This method bypasses the need for an explicit boundary action, allowing a direct algebraic characterization of the ASA. In that sense, our work aligns more closely with generalized bulk--based symmetry analyses, such as our earlier \(\mathfrak{sl}(3,\mathbb{R})\) study \cite{Ozer:2025bpb}, and represents a supersymmetric extension of that framework. We do not aim to construct a complete dual quantum mechanical theory; rather, we focus on the intrinsic gauge--theoretic origin of boundary symmetries in the supersymmetric BF setting.

Unlike earlier JT models based on \(\mathfrak{sl}(2,\mathbb{R})\), where symmetry breaking typically appears through boundary dilaton constraints or horizon regularity conditions, our formulation realizes symmetry reduction algebraically via residual bulk supersymmetry, without imposing explicit dynamical constraints or thermodynamic inputs. On the physical level, symmetry breaking happens dynamically due to the time-dependent dilaton supermultiplet in the bulk, not because of boundary thermodynamics.

The structure of this paper is as follows: Section~\ref{sec:quantumN1} introduces the quantum $\mathcal{N}=1$ superconformal algebra. It begins with an overview of the conformal structure and continues with the construction of its extended version incorporating conformal spins $(\Delta, 1{-}\Delta)$. Section~\ref{sec:sec21} develops the holographic dictionary for $\mathfrak{sl}(2,\mathbb{R})$ JT dilaton gravity. It starts from the BF theory formulation and elaborates on its boundary realizations, including both affine and conformal boundary conditions. Section~\ref{sec:osp} focuses on the $\mathcal{N}=1$ $\mathfrak{osp}(1|2)$ extension of higher--spin dilaton gravity. It analyzes the ASA under affine boundary conditions and then presents the superconformal boundary structure that emerges in the Drinfeld--Sokolov gauge. Section~\ref{sec:conclusion} summarizes the main results and discusses their theoretical and holographic significance. Section~\ref{sec:ack} acknowledges the contributors and funding sources. Finally, Appendix~\ref{sec:appa} provides a summary of notation, while Appendix~\ref{sec:appB} reviews the boundary theories and their associated symmetry structures.
\section{The quantum $\cN=1$ superconformal algebra}\label{sec:quantumN1}

We begin by summarizing the key structural elements of the quantum $\cN=1$ superconformal algebra, focusing on the aspects most relevant to the framework developed in this work.\,Rather than presenting a complete treatment, our goal is to emphasize the defining features that inform the supersymmetric extensions we explore.\,Originally formulated by A.B.\ Zamolodchikov~\cite{Zamolodchikov:1985wn}, the quantum $\cN=1$ superconformal algebra introduces a non--linear generalization of the Virasoro algebra that naturally accommodates higher--spin currents.
\subsection{Conformal structure of the quantum $\cN=1$ superconformal algebra}\label{sec21}

The algebra is built upon two primary fields: a fermionic current $G(z)$ of spin-$\tfrac{3}{2}$ and a bosonic stress--energy tensor $T(z)$ of spin-$2$. These fields admit standard mode expansions, given by $T(z) = \sum_n L_n z^{-n-2}$ and $G(z) = \sum_n G_n z^{-n-3/2}$. The essential structure of the quantum $\cN=1$ superconformal algebra is encoded in their non--trivial operator product expansions, which reflect the characteristic non--linearity of this supersymmetric extension of the Virasoro algebra.
\begin{align}
T(z_1) T(z_2) &\sim \frac{ \frac{c}{2}}{z_{12}^4} + \frac{2 T}{z_{12}^2} + \frac{\partial T}{z_{12}}, \\[10pt]
T(z_1) G(z_2) &\sim \frac{\frac{3}{2} G}{z_{12}^2} + \frac{\partial G}{z_{12}}, \\
G(z_1) G(z_2) &\sim  \frac{\frac{2c}{3}}{z_{12}^3} + \frac{2 T}{z_{12}}.
\end{align}
\subsection{Extended $\cN=1$ superconformal algebra with conformal spin--$(\Delta,1-\Delta)$  }
\vspace{0.5cm}
In this section, we construct an extended version of the quantum $\cN=1$ superconformal algebra by supplementing the conformal fields of spin $\Delta = 2$ and $\Delta = 3/2$, introduced in the previous subsection, with two additional components. These new fields, labeled $X$ and $Y$, are interpreted as dilaton partners and carry conformal weights $1 - \Delta = -1$ and $-1/2$, respectively. Their standard mode expansions are given by $X(z) = \sum_n X_n z^{-n+1}$ and $F(z) = \sum_n F_n z^{-n+1/2}$. Using Thielemans’ Mathematica package~\cite{Thielemans:1991uw}, one obtains the non--vanishing operator product expansions that define the extended $\cN=1$ superconformal algebra in this formulation.
\begin{align}
T(z_1) T(z_2) &\sim \frac{ \frac{c}{2}}{z_{12}^4} + \frac{2 T}{z_{12}^2} + \frac{\partial T}{z_{12}},\label{eq:Q1} \\[10pt]
T(z_1) G(z_2) &\sim \frac{\frac{3}{2} G}{z_{12}^2} + \frac{\partial G}{z_{12}}, \\
G(z_1) G(z_2) &\sim  \frac{\frac{2c}{3}}{z_{12}^3} + \frac{2 T}{z_{12}}, 
\end{align}
\begin{align}
T(z_1) X(z_2) &\sim - \frac{X}{z_{12}^2} + \frac{\partial X}{z_{12}}, \\[10pt]
T(z_1) F(z_2) &\sim - \frac{\frac{1}{2} F}{z_{12}^2} + \frac{\partial F}{z_{12}}, \\[10pt]
G(z_1) X(z_2) &\sim -\frac{2 F}{{ z_{12}}} ,\\[10pt]
G(z_1) F(z_2) &\sim - \frac{ X}{{z_{12}^{2}}} + \frac{\partial X}{2z_{12}}.\label{eq:Q2}
\end{align}
This extended version of the quantum $\cN=1$ superconformal algebra will later be contrasted with its classical counterpart, which will be explicitly derived in Section~\ref{sec:osp}.
\section{The $\mathfrak{sl}(2,\mathbb{R})$ holographic dictionary}\label{sec:sec21}

In this section, we explore the formalism of two--dimensional dilaton gravity defined on an $AdS_2$ background. We begin by outlining the essential features of dilaton gravity that are instrumental for constructing the holographic dictionary. This is followed by a systematic derivation of the ASA within the $\mathfrak{sl}(2,\mathbb{R})$ framework. We then turn our attention to the role of external source terms and the resulting holographic Ward identities, which provide key insights into the structure of higher--spin black hole configurations.
\subsection{JT dilaton gravity as a BF theory in two dimensions}\label{sec2}

This section provides a concise overview of higher--spin gravity in $AdS_2$, formulated within the framework of dilaton gravity. We adopt a gauge-theoretic perspective, treating dilaton gravity as a non--abelian theory. Specifically, our analysis focuses on the formulation of $AdS_2$ gravity in terms of the underlying $\mathfrak{sl}(2,\mathbb{R})$ algebraic structure.
\subsection{Connection to $\mathfrak{sl}(2,\mathbb{R})$ BF theory}

Analogous to the computational benefits afforded by Chern--Simons theory in three--dimensional gravity~\cite{Achucarro:1986uwr,Witten:1988hc}, alternative formulations in two dimensions offer substantial simplifications. In this work, we briefly summarize the key aspects of such approaches, with particular emphasis on the JT model. Specifically, in two dimensions, dilaton gravity with a negative cosmological constant admits an equivalent description in terms of a gauge theory defined over a spacetime manifold $\cM$:
\begin{equation}
    S[\cX,\cA] = \frac{k}{4\pi}\int_{\cM} \mathfrak{tr}[\cX \cF] +S_{\rm bdy} ,
\end{equation}
Here, $\mathcal{A}$ denotes a one--form connection with corresponding field strength $\cF = \mathrm{d}\cA + \cA \wedge \cA$, and $k$ is a coupling constant. The dilaton field $\cX$ is treated as a scalar taking values in the gauge algebra. The spacetime manifold $\cM$ is assumed to have the topology of a cylinder, $\cM \cong S^1 \times \mathbb{R}_+$, where the radial coordinate $\rho$ ranges over $0 \leq \rho < \infty$. The Euclidean time coordinate $y$ is periodically identified as $y \sim y + \beta$, with $\beta$ denoting the inverse temperature. A boundary term $S_{\text{bdy}}$ is included to ensure a well--defined variational principle, enforcing appropriate conditions at the asymptotic boundary $\partial \mathcal{M}$ located at infinite radius. The notational framework adopted in this work, including the components of the gauge fields, the dilaton variables, and the boundary charges, is based on standard conventions widely used in the literature. For the sake of clarity and consistency, the principal symbols and their associated physical roles are summarized in Appendix~\ref{sec:appa}.

The gauge fields $\cA$ and $\cX$ both take values in the Lie algebra $\mathfrak{sl}(2,\mathbb{R})$, and the trace operation $\mathfrak{tr}$ defines the bilinear form used to contract generators. To describe the dynamics of dilaton gravity in this setting, we utilize the basis of $\mathfrak{sl}(2,\mathbb{R})$ generators $\Lt_i$ with indices $i = 0, \pm 1$, which obey the following commutation relations:
\begin{eqnarray}
    \left[\Lt_i, \Lt_j\right] &=& (i-j) \Lt_{i+j}.
\end{eqnarray}
To proceed with explicit computations, we adopt a matrix representation for the generators:
\begin{equation}
\Lt_{-1}=\begin{pmatrix}0 & 0\\
1 & 0
\end{pmatrix}\quad,\quad \Lt_{0}=\frac{1}{2}\begin{pmatrix}-1 & 0\\
0 & 1
\end{pmatrix}\quad,\quad \Lt_{1}=\begin{pmatrix}0 & -1\\
0 & 0
\end{pmatrix}\;.\label{PTT-sl(2,R)-MR}
\end{equation}
With this representation, the only non--vanishing components of the invariant bilinear form are given by $\mathfrak{tr}(\Lt_{\mp1} \Lt_{\pm1}) = -2$ and $\mathfrak{tr}(\Lt_0 \Lt_0) = -\tfrac{1}{2}$. The action remains invariant under the gauge transformations:
\begin{equation}\label{gauge transformation}
\delta_{\lambda} \cA =d\lambda + \left[\cA,\lambda \right],\qquad \qquad \delta_{\lambda} \cX =[\cX,\lambda],
\end{equation}
Here, $\lambda$ is a gauge parameter taking values in the $\mathfrak{sl}(2,\mathbb{R})$ Lie algebra. These gauge transformations define the residual symmetries of the theory that preserve the asymptotic structure of the gauge fields, and they also determine how boundary symmetries emerge from bulk gauge invariance within the BF formulation. These elements provide the foundation for identifying the ASA, which in the case of $\mathfrak{sl}(2,\mathbb{R})$ corresponds to the Virasoro algebra. The presence of the dilaton field plays a pivotal role in determining which symmetries are preserved and which are broken at the boundary.
The equations of motion are given by
\begin{equation} \label{eom}
\mathcal{F} =d\cA+\cA\wedge \cA=0,\qquad \qquad d \cX + \left[\cA,\cX\right]=0.
\end{equation}
More importantly, these gauge transformations determine the residual symmetries that persist after imposing the radial gauge near the asymptotic boundary of $\mathrm{AdS}_2$. Consequently, they provide the foundational structure for identifying the ASA. The inclusion of the dilaton field $X$ is especially significant in this context: it couples to the gauge connection at the boundary and introduces dynamical constraints that restrict the admissible gauge transformations. This interaction results in a partial breaking of the full infinite--dimensional affine symmetry algebra, such as $\mathfrak{sl}(2,\mathbb{R})_k$ , to its corresponding finite--dimensional subalgebra, namely $\mathfrak{sl}(2,\mathbb{R})$. In this light, equations~\eqref{gauge transformation} and~\eqref{eom} do more than define local gauge redundancy; they also encapsulate the physical mechanism through which symmetry breaking is dynamically induced by the dilaton field.

Adopting the radial gauge, the gauge connections in an asymptotically $\mathrm{AdS}_2$ spacetime can be expressed in the following factorized form:
\begin{eqnarray}
\label{ads31}
    \cA &=& b^{-1} a\left(t\right) b + b^{-1} \mathrm{d}b, \quad
    \cX =b^{-1} x\left(t\right) b,
\end{eqnarray}
Here, the group element $b(\rho)$, which is state--independent and depends only on the radial coordinate, is chosen to take the form:
\begin{eqnarray}
\label{gr}
    b(\rho) = e^{\rho \Lt_{0}},
\end{eqnarray}
This representation allows for a more general class of metrics that incorporate all $\mathfrak{sl}(2,\mathbb{R})$ charges. Importantly, as long as $\delta b = 0$, the specific choice of $b$ does not affect the structure of the asymptotic symmetries. This freedom enables a more versatile formulation of the metric and requires boundary conditions that preserve such generality in the gravitational context. Additionally, within the radial gauge, the reduced connection $a(t)$ is a one-form valued in the $\mathfrak{sl}(2,\mathbb{R})$ Lie algebra and is independent of the radial coordinate, i.e., $a(t) = a_t(t)\, \mathrm{d}t$.

Our analysis focuses on the affine boundary conditions relevant to $\mathfrak{sl}(2,\mathbb{R})$ dilaton gravity. In doing so, we make use of the methodology developed in~\cite{Grumiller:2016pqb,Ozer:2017dwk,Ozer:2019nkv,Ozer:2021wtx,Ozer:2024ovo} to systematically derive the associated ASA. Based on these techniques, the most general solution consistent with asymptotically $AdS_2$ geometries can be expressed in the following general metric form:
\begin{gather}\label{AdSLineElement}
	\extd s^2 = \extd\rho^2 + 2\,\mathcal{L}^{0} \extd\rho \extd\varphi + \Big(\left(e^{\rho}\mathcal{L}^{+}
    -e^{-\rho}\mathcal{L}^{-}\right)^{2}+\left(\mathcal{L}^{0}\right)^{2}\Big)\,\extd\varphi^2 ~,
\end{gather}
which bears resemblance to its well--known $AdS_3$ counterpart~\cite{Grumiller:2016pqb,Ozer:2017dwk,Ozer:2019nkv,Ozer:2021wtx,Ozer:2024ovo,Sammani:2025wtx}. The dilaton field in this setting takes the general form:
\begin{gather}\label{AdSDilaton}
	X = e^{\rho}\,\mathcal{X}^{+} + e^{-\rho}\,\mathcal{X}^{-} ~.
\end{gather}
Accordingly, it becomes essential to specify affine $\mathfrak{sl}(2,\mathbb{R})$ boundary conditions that preserve this generalized metric structure.
\subsection{Affine boundary conditions}\label{osp21}

The objective of this section is to formulate $\mathfrak{sl}(2,\mathbb{R})$ higher--spin $AdS_2$ dilaton gravity under affine boundary conditions. Our analysis proceeds by computing the ASA associated with the most general (i.e., least restrictive) set of boundary data. We begin by introducing the $a_t$ component of the gauge connection, which takes values in the $\mathfrak{sl}(2,\mathbb{R})$ Lie algebra:
\begin{eqnarray}\label{bouncond999}
a_t &=& \alpha_i \mathcal{L}^i \Lt_i,
\end{eqnarray}
where the coefficients satisfy $\alpha_{0} = -2\alpha_{\pm1} = \frac{4}{k}$. The functions $\mathcal{L}^i$ are state-dependent and interpreted as dynamical boundary charges. Similarly, the dilaton field $x$ is expanded as
\begin{eqnarray}\label{bouncond888}
x &=& \cX^i \Lt_i,
\end{eqnarray}
which defines an additional set of three state--dependent functions $\cX^i$. Our aim is to derive the resulting ASA under affine boundary conditions through a canonical analysis. For this purpose, we systematically investigate all gauge transformations of the form~\eqref{gauge transformation} that preserve the boundary data encoded in equations~\eqref{bouncond999} and~\eqref{bouncond888}.

At this stage, it is convenient to parameterize the gauge transformation in terms of the $\mathfrak{sl}(2,\mathbb{R})$ Lie algebra basis:
\begin{equation}\label{boundarycond2}
    \lambda = b^{-1}\left[\epsilon^i \Lt_i \right] b.
\end{equation}
The gauge parameter $\lambda$ thus contains three bosonic functions $\epsilon^i$, which are arbitrary functions defined on the boundary. We now focus on those gauge parameters that satisfy the transformation law~\eqref{gauge transformation}. The corresponding infinitesimal variations of the dynamical fields are given by:
 \begin{align} 
       2\delta_\lambda\mathcal{L}^{0}	&= -\frac{k}{4}\partial_t\epsilon^{0} - \mathcal{L}^{+1}\epsilon^{-1} + \mathcal{L}^{-1}\epsilon^{+1}, \label{transf21} \\ 
    \delta_\lambda\mathcal{L}^{\pm 1}	&= \frac{k}{2}\partial_t\epsilon^{\mp 1} + \mathcal{L}^{0}\epsilon^{\mp 1} \mp \mathcal{L}^{\mp 1}\epsilon^{0}, \label{transf22} \\ 
    \delta_\lambda\mathcal{X}^{0}	&= 2\big(\mathcal{X}^{+1}\epsilon^{-1} - \mathcal{X}^{-1}\epsilon^{+1}\big), \label{transf23} \\ 
    \delta_\lambda\mathcal{X}^{\pm 1}	&=\mp \mathcal{X}^{0}\epsilon^{\pm 1} \pm \mathcal{X}^{\pm 1}\epsilon^{0}. \label{transf24}
\end{align} 
As a final step in this construction, we define the canonical boundary charge $\mathcal{Q}_a[\lambda]$, which generates the gauge transformations of the connection field described in Eqs.~\eqref{transf21}–\eqref{transf22}. In parallel, we introduce the canonical boundary charge $\mathcal{Q}_x[\lambda]$, associated with the transformations of the dilaton field as specified in Eqs.~\eqref{transf23}–\eqref{transf24}. 

The infinitesimal variation of these charges, which encodes the structure of the ASA, is given by~\cite{Banados:1994tn}:
\begin{equation}\label{Qvar}
    \delta_\lambda \mathcal{Q}_a = \frac{k}{2\pi} \int \mathrm{d}t\; \mathfrak{tr} \left(\lambda \delta a_{t} \right), \quad
    \delta_\lambda \mathcal{Q}_x = \frac{k}{2\pi} \int \mathrm{d}t\; \mathfrak{tr} \left(\lambda \delta x_{} \right).
\end{equation}
By performing a functional integration of the variations, we obtain the explicit form of the canonical boundary charges:
\begin{equation}\label{boundaryco9}
    \mathcal{Q}_a[\lambda] = \int \mathrm{d}t\; \left( \mathcal{L}^{i} \epsilon^{-i} \right), \quad
    \mathcal{Q}_x[\lambda] = \int \mathrm{d}t\; \left( \mathcal{X}^{j} \epsilon^{-j} \right).
\end{equation}
Having established both the infinitesimal gauge transformations and the associated canonical boundary charges, we now derive the ASA using the standard procedure outlined in~\cite{Blagojevic:2002du}. The algebraic structure is encoded in the following relation:
\begin{equation}\label{Qvar2}
    \delta_{\lambda} \digamma = \{\digamma, \mathcal{Q}_{a,x}[\lambda]\}
\end{equation}
for any generic phase space functional $\digamma$. The ASA is thus generated by the boundary charges $\mathcal{L}^i$ and $\mathcal{X}^i$. Ultimately, the corresponding operator product algebra takes the form:
\begin{eqnarray}\label{ope22}
    \mathcal{L}^{i}(\tau_1)\mathcal{L}^{j}(\tau_2) & \sim & \frac{\frac{k}{2}\eta^{ij}}{\tau_{12}^{2}} + \frac{(i-j)}{\tau_{12}} \mathcal{L}^{i+j}\\
    \mathcal{L}^{i}(\tau_1)\mathcal{X}^{j}(\tau_2) & \sim & \frac{(2i+j)}{\tau_{12}} \mathcal{X}^{i+j},\\
\end{eqnarray}
where $\tau_{12} = \tau_1 - \tau_2$. The coefficients $\eta^{ij} = \mathfrak{tr}(\Lt_i \Lt_j)$ define the components of the invariant bilinear form in the fundamental representation of the $\mathfrak{sl}(2,\mathbb{R})$ Lie algebra. An alternative and more compact representation of the operator product algebra can be written using index-free notation as follows:
\begin{eqnarray}\label{ope11}
    \mathfrak{\mathcal{\mathfrak{J}}}^{A}(\tau_1)\mathfrak{\mathcal{\mathfrak{J}}}^{B}(\tau_2) & \sim &
    \frac{\frac{k}{2}\eta^{AB}}{\tau_{12}^{2}} + \frac{\mathfrak{\mathcal{\mathfrak{f}}}^{AB}_{~~~C}
    \mathfrak{\mathcal{\mathfrak{J}}}^{C}}{\tau_{12}}. 
\end{eqnarray}
Here, $\eta^{AB}$ denotes the trace matrix, and $\mathfrak{f}^{AB}_{~~C}$ are the structure constants of the underlying algebra, with indices running over $(A, B = 0, \pm1)$. In particular, one has $\mathfrak{f}^{ij}_{~~i+j} = (i - j)$. Accordingly, under the most relaxed set of boundary conditions for $\mathfrak{sl}(2,\mathbb{R})$ dilaton gravity, the resulting ASA is identified with a single copy of the affine $\mathfrak{sl}(2,\mathbb{R})_k$ algebra.

The time--dependent profile of the dilaton field at the affine boundary does not modify the affine $\mathfrak{sl}(2,\mathbb{R})_k$ algebra itself. In the standard BF formulation, the ASA is determined by gauge transformations of the $\mathfrak{sl}(2,\mathbb{R})$--valued connection $a_t$, and this algebraic structure remains intact.

However, once the $\mathfrak{sl}(2,\mathbb{R})$--valued dilaton field $x$ is treated as a dynamical boundary degree of freedom, the set of background--preserving residual gauge transformations becomes restricted. 
This does not deform the affine algebra, but it affects how the symmetry is realized on the boundary phase space.

Consequently, while the ASA remains a single copy of affine $\mathfrak{sl}(2,\mathbb{R})_k$, the boundary dynamics is governed by an enlarged phase space that includes additional dynamical modes 
associated with the dilaton. In this sense, the Schwarzian action alone does not exhaust the full set of boundary degrees of freedom in the affine regime. The dilaton thus plays a dual role: it determines the geometric background and restricts the admissible residual gauge transformations to those that preserve a given configuration, while the affine algebra itself remains unchanged.

\subsection{Conformal boundary conditions}\label{bhreduction1}

In this section, our goal is to analyze the ASA associated with Brown--Henneaux-type boundary conditions. This gauge choice is specifically designed to make the asymptotic symmetry structure manifest and to streamline the analysis of boundary dynamics. By decoupling the radial dependence, it effectively isolates the relevant asymptotic degrees of freedom, thereby facilitating the computation of conserved quantities and the identification of corresponding boundary theories.

This gauge plays a central role in maintaining the full asymptotic symmetry group and in determining the effective boundary action, particularly within the context of holographic duality. To proceed, we impose the Drinfeld--Sokolov highest weight gauge condition on the $\mathfrak{sl}(2,\mathbb{R})$-valued connection defined in Eq.~\eqref{bouncond999}, which further restricts the allowed functional form of the coefficients. Under this reduction, the fields are constrained as follows:
\begin{eqnarray}
\mathcal{L}^0 = 0, \quad \mathcal{L}^{-1} = \mathcal{L}, \quad \alpha_{+1}\mathcal{L}^{+1} = 1.
\end{eqnarray}
where $\alpha_{-1} = \alpha$. It is important to note that these conformal boundary conditions correspond to the well--established Brown--Henneaux boundary conditions originally formulated in~\cite{Brown:1986nw} for $AdS_2$ gravity. Inspired by analogous developments in the context of three--dimensional gravity~\cite{Perez:2016vqo,Cardenas:2021vwo}, we propose that the gauge connection and the dilaton field adopt the following specific structure:
\begin{align}\label{bouncondhf}
a_t &= \Lt_{+1} + \alpha\mathcal{L} \Lt_{-1},\\
x &= \cX \Lt_{1}
 -\cX'\Lt_{0}
+ \bigg(\alpha\mathcal{L}\cX
             +\frac{\cX ''}{2}
       \bigg)\Lt_{-1}.
\end{align}
Here, $\alpha$ acts as a scaling parameter whose exact value will be fixed at a later stage. We define two key functionals: the boundary charge $\mathcal{L}$ and the dilaton field $\cX$. With these constraints in place, we are now prepared to derive the conformal ASA. As dictated by the Drinfeld--Sokolov reduction, the residual gauge parameter $\lambda$ depends on four independent functions, with $\epsilon \equiv \epsilon^{+1}$ serving as the primary functional degree of freedom. The gauge parameter then takes the following form:
\begin{eqnarray}\label{ebc08}
\lambda &=& b^{-1}\left[
  \epsilon \Lt_{1}
 -\epsilon'\Lt_{0}
+ \bigg( \alpha\mathcal{L}\epsilon
             +\frac{\epsilon ''}{2}
       \bigg)\Lt_{-1}      
\right]b.
\end{eqnarray}
The dilaton field $x$ is defined in the same form as the gauge parameter $\lambda$ in order to maintain consistency with the equations of motion and compatibility with the asymptotic symmetry group. This structural alignment allows the dilaton to behave as a coadjoint element in the gauge theory and to act as a stabilizing agent for the residual symmetries. Moreover, this choice ensures the well-posedness of the variational principle and enables the consistent derivation of the conserved Casimir function $C$. Thus, defining $x$ in the same form as $\lambda$ is crucial for preserving the asymptotic symmetry group, ensuring the consistency of the theory, and facilitating the formulation of boundary dynamics in a holographic context.

By substituting the gauge parameter $\lambda$ into the field transformation laws~\eqref{gauge transformation}, the resulting infinitesimal gauge transformations are found to be:
\begin{eqnarray}
\delta_{\lambda}\mathcal{L} &=&
\frac{1}{2 \alpha}  \epsilon ^{'''}
+\epsilon  \mathcal{L}'+2 \mathcal{L} \epsilon', \label{ebc08666} \\
\delta_{\lambda}\cX &=&
\epsilon  \cX'- \epsilon' \cX. \label{ebc08667}
\end{eqnarray}
Under the imposed boundary conditions, the corresponding canonical charges $\mathcal{Q}_a[\lambda]$ and $\mathcal{Q}_x[\lambda]$ remain integrable, thereby allowing $\mathcal{L}$ and $\mathcal{X}$ to be interpreted as conserved quantities. These gauge transformations are instrumental in uncovering the structure of the ASA~\cite{Banados:1998pi}.

By analyzing the asymptotic symmetries, one can integrate the variations of the canonical boundary charges, $\delta_\lambda \mathcal{Q}_a$ as given in Eq.~\eqref{ebc08666} and $\delta_\lambda \mathcal{Q}_x$ in Eq.~\eqref{ebc08667}, to arrive at the following expressions:
\begin{equation}\label{boundaryco11111333333}
    \mathcal{Q}_a[\lambda] = \int \mathrm{d}\varphi\; \left(\mathcal{L} \epsilon \right), \quad
    \mathcal{Q}_x[\lambda] = \int \mathrm{d}\varphi\; \left(\mathcal{X} \epsilon \right).
\end{equation}
These canonical boundary charges naturally induce an operator product algebra at the asymptotic boundary. In contrast to the standard two--dimensional CFT framework, a one-dimensional CFT description can be obtained by effectively suppressing one of the coordinates. This is achieved by setting $z_2 = 0$ and identifying $z_1$ with the boundary coordinate $\tau$, or more simply by taking $z = \tau$:
\begin{eqnarray}\label{ope2}
  \mathcal{L}(\tau_1)\mathcal{L}(\tau_2) &\sim& \frac{{3k}}{{2\tau_{12}^{4}}} + \frac{2\mathcal{L}}{{\tau_{12}^{2}}} + \frac{\mathcal{L}'}{\tau_{12}},\\
   \cL(\tau_1)\cX(\tau_2) &\sim& - \frac{\cX}{{\tau_{12}^{2}}} + \frac{\cX'}{\tau_{12}}.
  \end{eqnarray}
From these relations, it follows that $\cX$ transforms as a vector under boundary diffeomorphisms, while $\mathcal{L}$ behaves as a spin-two density. Throughout this analysis, the explicit dependence on the gauge parameter $\lambda$ has been omitted, as the boundary conditions no longer require it. In effect, the residual symmetries associated with the connection $a_t$ and the dilaton $x$ are interpreted as diffeomorphisms on the boundary circle, generated by the vector field $\epsilon$. It is thus observed that the ASA of $\mathfrak{sl}(2,\mathbb{R})$ dilaton gravity with conformal boundary conditions is given by a single copy of the Virasoro algebra with central charge $c = 3k$.

The conformal boundary profile of the dilaton field does not modify the Virasoro algebra obtained under Drinfeld--Sokolov boundary conditions. In the conventional BF formulation, the ASA 
is generated by gauge transformations of the $\mathfrak{sl}(2,\mathbb{R})$-valued connection $a_t$, and this algebraic structure, including its central charge, remains intact. However, once the dilaton field $x$ is promoted to a dynamical boundary degree of freedom, the set of background-preserving residual gauge transformations becomes restricted. This restriction does not deform the Virasoro algebra itself, 
but it affects how the symmetry is realized on the boundary phase space.

Consequently, while the ASA remains a single copy of the Virasoro algebra with central charge $c=3k$, the boundary dynamics involves additional dilaton-dependent modes that are not fully captured by the Schwarzian action alone. In this sense, the dilaton plays a dual role: it determines the background geometry and constrains the admissible residual symmetries without altering 
the underlying algebraic structure.
\section{The $\mathcal{N}=1$ $\mathfrak{osp}(1|2)$ higher--spin dilaton supergravity}\label{sec:osp}

The analysis presented in this section is not in contradiction with the framework developed in Ref.~\cite{Cardenas:2018krd}, but rather aims to revisit closely related physical structures from a complementary conceptual and structural perspective. In particular, the notion of ``asymptotic symmetry breaking'' employed throughout this work should not be identified with the symmetry reduction arising from holonomy conditions, regularity requirements, or Casimir fixing in the sense discussed in Ref.~\cite{Cardenas:2018krd}. Instead, the emphasis here is placed on the restriction of admissible residual gauge transformations induced by the dynamical behavior of the dilaton supermultiplet.

In this sense, the term ``breaking'' does not refer to a violation of the underlying ASA, but to an algebraic and dynamical selection of stabilizers within the BF--theoretic formulation. As in Ref.~\cite{Cardenas:2018krd}, affine and superconformal boundary conditions are initially preserved. However, the focus of the present analysis lies on identifying a restrictive mechanism that already emerges at the level of bulk residual gauge symmetry, prior to imposing holonomy or additional regularity conditions. Accordingly, this section does not aim to reproduce the boundary dynamics derived in Ref.~\cite{Cardenas:2018krd}, but rather to elucidate the BF--theoretic origin of the associated asymptotic symmetry structure and to clarify the role played by the dilaton supermultiplet in shaping the physically admissible phase space.

Building upon the canonical analysis of the $\mathcal{N}=1$ $\mathfrak{osp}(1|2)$ case, we now formulate the two--dimensional higher-spin dilaton supergravity theory with negative cosmological constant in the BF framework. This theory can be equivalently described by the following action on a spacetime manifold $\cM$:
\begin{equation}
    S[\cX,\cA] = \frac{k}{4\pi} \int_{\cM} \mathfrak{str}[\cX \cF] + S_{\rm bdy},
\end{equation}
where $\mathcal{A}$ is a one--form superconnection with corresponding field strength $\cF = \mathrm{d} \cA + \cA \wedge \cA$, and $k$ denotes a coupling constant. The boundary term $S_{\rm bdy}$ is included to ensure a well-defined variational principle by enforcing suitable boundary conditions at $\partial \mathcal{M}$, which corresponds to the asymptotic boundary at infinite radius.

The dilaton field $\cX$ is an algebra-valued scalar taking values in the Lie superalgebra $\mathfrak{osp}(1|2)$, just like the superconnection $\cA$. The supertrace operation $\mathfrak{str}$ provides the invariant bilinear form used to contract the generators of the superalgebra.

To describe the structure of the dilaton supergravity system, we now introduce the underlying $\mathfrak{osp}(1|2)$ superalgebra. It consists of bosonic generators $L_n$ with $n = 0, \pm 1$ and fermionic generators $G_p$ with $p = \pm \tfrac{1}{2}$. These generators satisfy the following graded (anti-)commutation relations:
\begin{equation}
	\begin{split}
		\sbk{L_{m},L_{n}} &\,=\, (m-n)L_{m+n}  \ , \quad \sbk{L_{m},G_{p}} = \left(\frac{1}{2} m -p \right) G_{m+p} \ ,\\
		 \cbk{G_{p},G_{q}} &\,=\, - 2  L_{p+q} \ .
	\end{split}
\end{equation}
An explicit matrix representation of the $\mathfrak{osp}(1|2)$ generators can be written as:
\begin{equation}
	\begin{split}
		L_{1} \,=\,& \left[ \begin{array}{c c|c}
	0 & 0 & 0 \\
	1 & 0 & 0 \\
	\hline
	0 & 0 & 0 \\
\end{array} \right] \ , \quad L_{0} \,=\,\left[ \begin{array}{cc|c}
			\frac{1}{2} & 0 & 0\\
			0 & - \frac{1}{2} & 0  \\   
			\hline 
			0& 0 & 0   \\
		\end{array}\right] 
	 \ , \quad 
L_{-1} \,=\, \left[ \begin{array}{c c|c}
	0 & -1 & 0 \\
	0 & 0 & 0 \\
	\hline
	0 & 0 & 0 \\
\end{array} \right]  \ ,\\
G_{\frac{1}{2}} \,=\,& \left[ \begin{array}{c c|c}
0 & 0 & 0 \\
0 & 0& 1 \\
\hline
-1 & 0 & 0\\
\end{array}\right] \ , \quad G_{-\frac{1}{2}} \,=\, \left[ \begin{array}{c c|c}
0 & 0 &1 \\
0 & 0& 0 \\
\hline
0 & 1 & 0\\
\end{array}\right] \ .
	\end{split}
\end{equation}
In this basis, the invariant bilinear form is defined via the supertrace, and is given by:
\begin{equation}
    \mathfrak{str} \big(L_n,L_m\big) =
    \begin{pmatrix}
    0 & 0 & -1\\
    0 & \frac{1}{2} & 0\\
    -1 & 0 & 0
    \end{pmatrix}, \quad
   \mathfrak{str} \cbk{G_{p}G_{q}}=
    \begin{pmatrix}
    0 & 2\\
    -2 & 0
    \end{pmatrix}.
\end{equation}
where the supertrace $\mathfrak{str}$ of a matrix $\Mt$ is defined as
\begin{equation}
    \mathfrak{str} \Mt \equiv \Mt_{11} + \Mt_{22} - \Mt_{33}.
\end{equation} 
Section~\ref{sec:affinee} defines a regime in which no highest-weight constraints 
are imposed on the state-dependent functions. The resulting asymptotic symmetry 
algebra is the full $\mathcal{N}=1$ affine $\mathfrak{osp}(1|2)_k$ algebra. 
In contrast, Section~\ref{sec:bhreduction2} introduces Drinfeld--Sokolov boundary 
conditions, which explicitly implement highest-weight constraints and thereby 
reduce the affine algebra to the classical $\mathcal{N}=1$ superconformal algebra. 
These two regimes correspond to distinct realizations of the boundary phase space 
and its associated symmetry structure.
\subsection{$\mathcal{N}=1$ affine boundary conditions}\label{sec:affinee}
\noindent
The affine boundary conditions considered in this subsection are the supersymmetric counterpart of the bosonic affine setup discussed in Section~\ref{osp21} for $\mathfrak{sl}(2,\mathbb{R})$ JT dilaton gravity. In particular, the asymptotic symmetry is initially described by the affine $\mathfrak{osp}(1|2)_k$ algebra, while the dilaton supermultiplet dynamically restricts the set of admissible background-preserving residual gauge transformations.

The goal of this section is to construct an extended higher--spin $AdS_2$ framework by formulating it as an $\mathcal{N}=1$ $\mathfrak{osp}(1|2)$ dilaton supergravity theory defined with affine boundary conditions. Our analysis aims to determine the resulting ASA under the most general boundary conditions consistent with this setup. 

Following the discussion in the previous section, we adopt the principal embedding of $\mathfrak{sl}(2,\mathbb{R})$ into $\mathfrak{osp}(1|2)$ as a subalgebra. This embedding provides a natural foundation for implementing affine boundary conditions in asymptotically $AdS_2$ spacetimes. To proceed with this construction, we introduce the gauge connection in the following form:
\begin{align}\label{bouncondaf}
    a_t = \alpha_i\mathcal{L}^i \Lt_{i}
    +\beta_i\mathcal{G}^i \Gt_{i}
   \end{align}
where the scaling relations among the parameters are given by $\alpha_0 = -2\alpha_{\pm1}=\mp4 \beta_{\pm\frac{1}{2}}$. As a consequence, the gauge connection contains five independent dynamical functions, denoted by $\mathcal{L}^i$ and $\mathcal{G}^i$, which are interpreted as the corresponding boundary charges.

The dilaton field $x$ is expanded in the same $\mathfrak{osp}(1|2)$ basis and takes the form:
\begin{eqnarray}\label{bouncond888}
x &=&\mathcal{X}^i \Lt_{i}+ \mathcal{Y}^i \Gt_{i}
\end{eqnarray}
In this framework, we obtain eight additional state-dependent functions, $\cX^i$ and $\cY^i$, arising from the dilaton field expansion. Our goal is to derive the ASA associated with these affine boundary conditions via a canonical analysis. To this end, we systematically investigate all gauge transformations that preserve the structure of the boundary conditions defined above.

At this stage, it is convenient to express the gauge parameter in terms of the $\mathfrak{osp}(1|2)$ Lie superalgebra basis:
\begin{equation}\label{boundarycond2}
    \lambda = b^{-1}\left[\epsilon^i \Lt_{i}+\eta^i \Gt_{i} \right]b.
\end{equation}
Here, the gauge parameter consists of five bosonic functions, $\epsilon^i$ and $\eta^i$, which are arbitrary functions defined on the boundary. We now proceed to analyze those gauge parameters that satisfy the transformation law~\eqref{gauge transformation}, and determine the corresponding infinitesimal gauge transformations for both the connection and the dilaton fields.
\begin{align}
    \delta_\lambda \mathcal{L}^{\mp 1} &= -\frac{k}{2} \partial_t \epsilon^{\mp 1} \pm \eta^{\mp \frac{1}{2}} \mathcal{G}^{\mp \frac{1}{2}} \mp 2 \mathcal{L}^0 \epsilon^{\mp 1} \mp \mathcal{L}^{\mp 1} \epsilon^0, \label{eq:deltaLpm1} \\
    \delta_\lambda \mathcal{L}^0 &= \frac{k}{4} \partial_t \epsilon^0 - \frac{1}{2} \eta^{\frac{1}{2}} \mathcal{G}^{-\frac{1}{2}} + \frac{1}{2} \eta^{-\frac{1}{2}} \mathcal{G}^{\frac{1}{2}} - \mathcal{L}^1 \epsilon^{-1} + \mathcal{L}^{-1} \epsilon^1, \label{eq:deltaL0} \\
    \delta_\lambda \mathcal{G}^{\mp \frac{1}{2}} &= \pm k \partial_t \eta^{\mp \frac{1}{2}} + 2 \eta^{\pm \frac{1}{2}} \mathcal{L}^{\mp 1} + 2 \eta^{\mp \frac{1}{2}} \mathcal{L}^0 - \mathcal{G}^{\pm \frac{1}{2}} \epsilon^{\mp 1} \mp \frac{1}{2} \mathcal{G}^{\mp \frac{1}{2}} \epsilon^0, \label{eq:deltaGpm12} \\
    \delta_\lambda \mathcal{X}^{\mp 1} &= -2 \eta^{\mp \frac{1}{2}} \mathcal{Y}^{\mp \frac{1}{2}} \mp \mathcal{X}^{\mp 1} \epsilon^0 \pm \mathcal{X}^0 \epsilon^{\mp 1}, \label{eq:deltaXpm1} \\
    \delta_\lambda \mathcal{X}^0 &= -2 \eta^{\frac{1}{2}} \mathcal{Y}^{-\frac{1}{2}} - 2 \eta^{-\frac{1}{2}} \mathcal{Y}^{\frac{1}{2}} + 2 \mathcal{X}^1 \epsilon^{-1} - 2 \mathcal{X}^{-1} \epsilon^1, \label{eq:deltaX0} \\
    \delta_\lambda \mathcal{Y}^{\mp \frac{1}{2}} &= \pm \frac{1}{2} \eta^{\mp \frac{1}{2}} \mathcal{X}^0 \mp \eta^{\pm \frac{1}{2}} \mathcal{X}^{\mp 1} \pm \mathcal{Y}^{\pm \frac{1}{2}} \epsilon^{\mp 1} \mp \frac{1}{2} \mathcal{Y}^{\mp \frac{1}{2}} \epsilon^0. \label{eq:deltaYpm12}
\end{align}
As a final step in this construction, we define the canonical boundary charge $\mathcal{Q}_{a,x}[\lambda]$, which generates the gauge transformations presented in Eqs.~\eqref{eq:deltaLpm1}--\eqref{eq:deltaYpm12}. The variation of this charge, which encodes the structure of the ASA, is given by~\cite{Banados:1994tn}:
\begin{equation}\label{Qvar}
    \delta_\lambda \mathcal{Q}_a = \frac{k}{2\pi} \int \mathrm{d}t\; \mathfrak{tr} \left(\lambda \delta a_{t} \right), \quad
    \delta_\lambda \mathcal{Q}_x = \frac{k}{2\pi} \int \mathrm{d}t\; \mathfrak{tr} \left(\lambda \delta x_{} \right).
\end{equation}
Integrating this variation functionally yields the explicit expression for the canonical boundary charges:
\begin{equation}\label{boundaryco9}
    \mathcal{Q}_a[\lambda] = \int \mathrm{d}t\; \left( \mathcal{L}^{i} \epsilon^{-i}+\mathcal{G}^{i}\eta^{-i} \right), \quad
    \mathcal{Q}_x[\lambda] = \int \mathrm{d}t\; \left( \mathcal{X}^{i} \epsilon^{-i}+\mathcal{Y}^{i}\eta^{-i} \right).
\end{equation}
Having established both the infinitesimal gauge transformations and the canonical boundary charge, we are now in a position to derive the ASA using the standard procedure outlined in~\cite{Blagojevic:2002du}. This is accomplished through the fundamental relation:
\begin{equation}\label{Qvar2}
    \delta_{\lambda} \digamma = \{\digamma, \mathcal{Q}_{a,x}[\lambda]\}
\end{equation}
for an arbitrary phase space functional $\digamma$. In this setting, the charges $\mathcal{L}^i$ and $\mathcal{G}^i$, along with their corresponding dilaton partners $\mathcal{X}^i$ and $\mathcal{Y}^i$, serve as the generators of the ASA. Ultimately, the operator product algebra takes the following form:
\begin{eqnarray}\label{ope22}
\mathcal{L}^{i}(\tau_1)\mathcal{L}^{j}(\tau_2) & \sim & \frac{\frac{k}{2}\eta^{ij}}{\tau_{12}^{2}} + \frac{(i-j)}{\tau_{12}} \mathcal{L}^{i+j},\\
\mathcal{L}^{i}(\tau_1)\mathcal{G}^{j}(\tau_2) & \sim &                                              \frac{(\frac{i}{2}-j)}{\tau_{12}} \mathcal{G}^{i+j},\\
\mathcal{G}^{i}(\tau_1)\mathcal{G}^{j}(\tau_2) & \sim & \frac{\frac{k}{2}\theta^{ij}}{\tau_{12}^{2}} + \frac{2}{\tau_{12}} \mathcal{L}^{i+j},\\
\mathcal{L}^{i}(\tau_1)\mathcal{X}^{j}(\tau_2) & \sim & \frac{(2i+j)}{\tau_{12}} \mathcal{X}^{i+j},\\
\mathcal{L}^{i}(\tau_1)\mathcal{Y}^{j}(\tau_2) & \sim & \frac{(\frac{i}{2}+j)}{\tau_{12}} \mathcal{Y}^{i+j},\\
\mathcal{G}^{i}(\tau_1)\mathcal{X}^{j}(\tau_2) & \sim &  \frac{2}{\tau_{12}} \mathcal{Y}^{i+j},\\
\mathcal{G}^{i}(\tau_1)\mathcal{Y}^{j}(\tau_2) & \sim &  \frac{ (\frac{3i}{2}+\frac{j}{2})}{\tau_{12}} \mathcal{Y}^{i+j}.
\end{eqnarray}
where $\tau_{12} = \tau_1 - \tau_2$. In this formulation, $\eta^{ij} = \mathfrak{tr}(\Lt_i \Lt_j)$ and $\theta^{ij} = \mathfrak{tr}(\Gt_i \Gt_j)$ denote the invariant bilinear forms in the fundamental representation of the $\mathfrak{osp}(1|2)$ Lie superalgebra. The operator product algebra may be recast in a more compact, index-free form as follows:
\begin{eqnarray}\label{ope11}
    \mathfrak{\mathcal{\mathfrak{J}}}^{A}(\tau_1)\mathfrak{\mathcal{\mathfrak{J}}}^{B}(\tau_2) & \sim &
    \frac{\frac{k}{2}\eta^{AB}}{\tau_{12}^{2}} + \frac{\mathfrak{\mathcal{\mathfrak{f}}}^{AB}_{~~~C}
    \mathfrak{\mathcal{\mathfrak{J}}}^{C}}{\tau_{12}}. 
\end{eqnarray}
Here, $\eta^{AB}$ denotes the invariant trace matrix, while $\mathfrak{f}^{AB}_{~~C}$ represent the structure constants of the $\mathfrak{osp}(1|2)$ algebra, with indices $(A,B = 0, \pm1, \pm\tfrac{1}{2})$. These constants satisfy the relation $\mathfrak{f}^{ij}_{~~i+j} = (i - j)$. Under the most general (least restrictive) boundary conditions, the ASA of $\mathfrak{osp}(1|2)$ dilaton supergravity is characterized by a single copy of the affine $\mathfrak{osp}(1|2)_k$ algebra.

\noindent In the context of $\mathfrak{osp}(1|2)$ JT dilaton supergravity with affine boundary conditions, the ASA is given by a single copy of the infinite--dimensional affine $\mathfrak{osp}(1|2)_k$ algebra. The affine $\mathfrak{osp}(1|2)_k$ algebra remains unchanged.

The role of the dilaton supermultiplet is not to deform the affine algebra itself, but to influence the realization of its symmetries on the boundary phase space. Once a specific time--dependent dilaton background is specified, only those residual gauge transformations that preserve this background remain admissible. In other words, the set of realized asymptotic symmetries is dynamically reduced to the stabilizer subalgebra of the chosen dilaton configuration.

In this work, the term ``asymptotic symmetry breaking'' refers precisely to this reduction of admissible residual gauge transformations. It does not signify a deformation of the affine $\mathfrak{osp}(1|2)_k$ algebra, nor a modification of its structure constants or central extension. The realized symmetry is determined by the background-preserving residual gauge transformations selected by the dynamical dilaton supermultiplet.

This mechanism is structurally analogous to the symmetry reduction observed in the bosonic $\mathfrak{sl}(3,\mathbb{R})$ extension of JT gravity~\cite{Ozer:2025bpb}, where the affine algebra remains intact but the realized symmetry is reduced to the stabilizer subalgebra selected by the dilaton background.

The symmetry reduction originates from constraints imposed by the dilaton on the boundary phase space. Consequently, the residual symmetries encoded in the fields $a_t$ and $x$ correspond to boundary diffeomorphisms generated, respectively, by the bosonic and fermionic components $\epsilon^i$ and $\eta^i$ of the gauge parameter.
\subsection{$\mathcal{N}=1$ superconformal boundary conditions}\label{sec:bhreduction2}

The superconformal boundary conditions imposed in this subsection provide the supersymmetric analogue of the bosonic superconformal (Drinfeld--Sokolov) reduction discussed in Section~\ref{bhreduction1}. Accordingly, the emergence of the classical $\mathcal{N}=1$ superconformal algebra originates from the imposed highest-weight (DS) constraints on the boundary data, rather than from a dilaton-induced dynamical selection.

In contrast to the affine regime discussed in Section~\ref{sec:affinee}, we now impose Drinfeld--Sokolov (highest weight) boundary conditions on the $\mathfrak{osp}(1|2)$-valued connection.
This gauge choice implements an explicit algebraic reduction of the affine $\mathfrak{osp}(1|2)_k$ symmetry to the classical $\mathcal{N}=1$ superconformal algebra.

Unlike the dynamical symmetry restriction induced by the dilaton supermultiplet in the affine case, the present reduction originates directly from the boundary constraints themselves.
Consequently, the emergence of the super-Virasoro structure is a structural consequence of the imposed highest-weight gauge.

In this section, we aim to analyze the ASA associated with the Brown--Henneaux boundary conditions in the context of $\mathfrak{osp}(1|2)$ supergravity. To this end, we impose the Drinfeld--Sokolov highest weight gauge condition on the $\mathfrak{osp}(1|2)$-valued connection defined in Eq.~\eqref{bouncondaf}, which further restricts the coefficient structure. This gauge fixing yields the following field constraints:
\begin{equation}
\mathcal{L}^{0} = \mathcal{G}^{+\frac{1}{2}} = 0, \quad 
\mathcal{L}^{-1} = \mathcal{L}, \quad 
\mathcal{G}^{-\frac{1}{2}} = \mathcal{G}, \quad 
\alpha_{+} \mathcal{L}^{+1} = 1,
\end{equation}
and introduces the new scaling parameters $\alpha_{-1} = \alpha$ and $\beta_{-\frac{1}{2}} = \beta$.

It is worth noting that these conformal boundary conditions are in direct analogy with the well-known Brown--Henneaux boundary conditions initially developed in~\cite{Brown:1986nw} for $AdS_3$ gravity. Our approach draws on boundary conditions previously explored in the three--dimensional setting~\cite{Perez:2016vqo,Cardenas:2021vwo}. In this spirit, we propose that the gauge connection and the dilaton field take the following form:
\begin{align}\label{bouncondhf}
a_t =&\Lt_{+1} + \alpha\mathcal{L} \Lt_{-1} +\beta\mathcal{G} \Gt_{-\frac{1}{2}},\\
x =& \mathcal{X}\Lt_1+\mathcal{Y}  \Gt_{\frac{1}{2}}- \mathcal{X} '\Lt_0+ \left(\alpha  \mathcal{G} \mathcal{X} -\mathcal{Y} '\right)\Gt_{-\frac{1}{2}}\\
+&  \left( \alpha  \mathcal{Y}  \mathcal{G}+ \beta  \mathcal{L} \mathcal{X} +\frac{1}{2}\mathcal{X} ''\right)\Lt_{-1}
\end{align}
After implementing the above procedures, we are now positioned to derive the conformal ASA. As a consequence of the Drinfeld--Sokolov reduction, the residual gauge parameter $\lambda$ is effectively described by only four independent functions. These are denoted by $\epsilon \equiv \epsilon^{+1}$ and $\eta \equiv \eta^{+\frac{1}{2}}$, and are explicitly given by:
\begin{align}\label{ebc08}
\lambda =
&b^{-1}\Bigg[\epsilon\Lt_1+\eta  \Gt_{\frac{1}{2}}- \epsilon '\Lt_0+ \left(\alpha  \mathcal{G} \epsilon -\eta '\right)\Gt_{-\frac{1}{2}}\\
+&  \left( \alpha  \eta  \mathcal{G}+ \beta  \mathcal{L} \epsilon +\frac{1}{2}\epsilon ''\right)\Lt_{-1}\Bigg]b.\nonumber
\end{align}
By substituting this gauge parameter into the field transformation equation given in Eq.~\eqref{gauge transformation}, we obtain the following infinitesimal gauge transformations:
\begin{align}\label{ebc086}
\delta_{\lambda}\mathcal{L}&=\frac{k\epsilon^{(3)}}{4}+2\mathcal{L}\epsilon'+\epsilon\mathcal{L}'-\frac{1}{2}\eta\mathcal{G}'-\frac{3}{2}\mathcal{G}\eta'\\
\delta_{\lambda}\mathcal{G}&=\frac{c \eta ''}{3}+\epsilon  \mathcal{G}'+\frac{3 \mathcal{G} \epsilon '}{2}+2 \eta  \mathcal{L},\\
\delta_{\lambda}\cX&=\epsilon  \mathcal{X}'-\mathcal{X} \epsilon '-2 \eta  \mathcal{Y},\\
\delta_{\lambda}\cY&=\frac{\eta  \mathcal{X}'}{2}-\mathcal{X} \eta '+\epsilon  \mathcal{Y}'-\frac{\mathcal{Y} \epsilon '}{2}
\end{align}
Under the specified boundary conditions, the canonical boundary charges $\mathcal{Q}_a[\lambda]$ are both well-defined and integrable, allowing the fields $\mathcal{L}$ and $\mathcal{G}$ to be interpreted as conserved charges. Likewise, the charges $\mathcal{Q}_x[\lambda]$ associated with the dilaton sector are also integrable, enabling a similar interpretation for $\mathcal{X}$ and $\mathcal{Y}$.

Importantly, these gauge transformations provide key insights into the structure of the ASA~\cite{Banados:1998pi}. By analyzing the behavior of these asymptotic symmetries, one can determine the integral form of the variation of the canonical boundary charges,
\begin{equation}\label{boundaryco11111333333}
    \mathcal{Q}_a[\lambda] = \int \mathrm{d}\varphi\; \left(\mathcal{L} \epsilon +\mathcal{G}\eta \right), \quad
    \mathcal{Q}_x[\lambda] = \int \mathrm{d}\varphi\; \left(\mathcal{X} \epsilon +\mathcal{Y}\eta\right).
\end{equation}
These canonical boundary charges serve as a fundamental tool in characterizing the asymptotic operator product expansion at the conformal boundary:
\begin{align}
 \mathcal{L}(\tau_1)\mathcal{L}(\tau_2) \sim& \frac{{\frac{3k}{2}}}{{\tau_{12}^{4}}} + \frac{2\mathcal{L}}{{\tau_{12}^{2}}} + \frac{\mathcal{L}'}{\tau_{12}},\label{eq:ope22}\\
\mathcal{L}(\tau_1)\mathcal{G}(\tau_2) \sim&   \frac{\frac{3}{2}\mathcal{G}}{{\tau_{12}^{2}}} + \frac{\mathcal{G}'}{\tau_{12}},\label{eq:ope23}\\
\mathcal{G}(\tau_1)\mathcal{G}(\tau_2) \sim& \frac{{2 k}}{{\tau_{12}^{3}}} + \frac{2\mathcal{L}}{\tau_{12}}. \label{eq:ope3232}
\end{align}
These operator product expansions define the classical $\mathcal{N}=1$ superconformal algebra, which extends the Virasoro algebra through the inclusion of the spin-$\tfrac{3}{2}$ current $\mathcal{G}$. In particular, the operator product~\eqref{eq:ope23} captures the interaction between the spin--2 and spin--$\tfrac{3}{2}$ modes.

With regard to the dilaton sector, the operator products involving $\mathcal{X}$ and $\mathcal{Y}$ reveal additional structural features:
\begin{align}
\cL(\tau_1)\cX(\tau_2) \sim& - \frac{\cX}{{\tau_{12}^{2}}} + \frac{\cX'}{\tau_{12}},\\
\cL(\tau_1)\cY(\tau_2) \sim& - \frac{\frac{1}{2}\cY}{{\tau_{12}^{2}}} +\frac{\cY'}{\tau_{12}},\\
\mathcal{G}(\tau_1)\cX(\tau_2) \sim& - \frac{2\cY}{{\tau_{12}^{2}}},\\
\cG(\tau_1)\cY(\tau_2) \sim& - \frac{\cX}{{\tau_{12}^{2}}} + \frac{\cX'}{2\tau_{12}}.
  \end{align}
These structures provide a natural framework for organizing the asymptotic symmetries in $\mathcal{N}=1$ $\mathfrak{osp}(1|2)$ dilaton supergravity and play a central role in determining the boundary dynamics in the presence of symmetry breaking. While the relations described above correspond to the undeformed $\mathcal{N}=1$ superconformal algebra, the inclusion of dilaton fields $\cX$ and $\cY$ in the extended $\mathfrak{osp}(1|2)$ model modifies this structure. These fields contribute additional terms to the algebra through boundary constraints, resulting in symmetry breaking and a deformation of the standard commutation relations. The deformed structure retains the essential features of the $\mathcal{N}=1$ superconformal algebra, while also encoding the dynamical influence of the dilaton sector.

From a physical perspective, the $\mathcal{N}=1$ superconformal algebra governs the dynamics of spin--$2$ and spin--$\tfrac{3}{2}$ boundary currents. The generator $\mathcal{L}$ corresponds to the energy--momentum tensor, while $\mathcal{G}$ represents a conserved higher--spin current of conformal weight--$\tfrac{3}{2}$. In the holographic setting, nontrivial $\mathcal{N}=1$ superconformal charges reflect the presence of spin--$\tfrac{3}{2}$ hair at the $\mathrm{AdS}_2$ boundary, leading to higher--derivative deformations in the dual boundary theory. The incorporation of the dilaton fields $\mathcal{X}$ and $\mathcal{Y}$ modifies the standard $\mathcal{N}=1$ algebra by introducing symmetry-breaking terms, thereby revealing how higher--spin symmetries become partially broken due to boundary conditions. This deformation parallels the symmetry-breaking mechanism observed in the Schwarzian limit of $\mathfrak{sl}(2,\mathbb{R})$ JT gravity, now generalized to spin-$\tfrac{3}{2}$ dynamics.

It follows that $\mathcal{X}$ and $\mathcal{Y}$ behave as boundary vectors, while $\mathcal{L}$ and $\mathcal{G}$ function as spin--$2$ and spin--$\tfrac{3}{2}$ densities, respectively. In this work, we have reformulated the boundary conditions in such a way that eliminates any explicit dependence on the gauge parameter $\lambda$. Consequently, the residual symmetries associated with $a_t$ and $x$ correspond to circle diffeomorphisms generated by the vector fields $\epsilon$ and $\eta$. For the conformal boundary conditions in $\mathfrak{osp}(1|2)$ dilaton supergravity, the resulting ASA is identified as a single copy of the $\mathcal{N}=1$ $\mathfrak{osp}(1|2)$ superconformal algebra with central charge $c = 3k$.
\begin{table}[ht]
    \centering
    \begin{tabular}{|c|c||c|c|}
        \hline
        charge  & \textbf{$\Delta$} &  dilaton  & \textbf{1$-\Delta$} \\
        \hline
         $\cL$ & 2 & $\cX$ & $-1$  \\
         \hline
         $\cG$ & $\frac{3}{2}$ & $\cY$ & $-\frac{1}{2}$  \\  
         \hline
    \end{tabular}
    \caption{superconformal spins for superconformal charges and related dilatons.}
    \label{tab:symbols}
\end{table}
 In the context of $\mathfrak{sl}(2,\mathbb{R})$ JT dilaton gravity, as well as in its $\mathfrak{osp}(1|2)$ extension under conformal boundary conditions, the ASA at the boundary is initially identified with the infinite--dimensional $\mathcal{N}=1$ superconformal algebra. However, the presence of a time--dependent dilaton field alters the algebraic structure of the asymptotic symmetry. As a result, even the expected preservation of the finite $\tt{O} \tt{S} p(1|2)$  subgroup within the full $\mathcal{N}=1$ $\mathfrak{osp}(1|2)$ symmetry is dynamically reduced, and the remaining infinite--dimensional symmetry components are broken.
  
 \medskip

\noindent
To illustrate the mechanism of symmetry breaking induced by the dilaton supermultiplet, we present a simplified toy model based on a static, partially truncated configuration. Consider the following profile for the $\mathfrak{osp}(1|2)$--valued dilaton field:
\[
\mathcal{X}^0 = \text{const}, \quad \mathcal{X}^{\pm 1} = 0, \quad \mathcal{Y}^{\pm \frac{1}{2}} = 0.
\]
This choice retains only the $\mathcal{X}^0$ component while suppressing all fermionic modes $\mathcal{Y}^{\pm \frac{1}{2}}$ and the bosonic charges $\mathcal{X}^{\pm 1}$. Under the gauge transformations generated by the full affine algebra (Eqs.~\eqref{eq:deltaLpm1}--\eqref{eq:deltaYpm12}),  one finds that such a configuration leads to a restricted set of residual symmetries. In particular, the variations $\delta_\lambda \mathcal{Y}^{\pm \frac{1}{2}}$ and $\delta_\lambda \mathcal{X}^{\pm 1}$ vanish identically, while $\delta_\lambda \mathcal{X}^0$ survives only for $\epsilon^0$ and a subset of $\eta$--parameters. As a consequence, the boundary charges associated with broken modes cease to transform non--trivially and are effectively projected out of the ASA. This demonstrates, at a minimal level, how specific dilaton backgrounds can dynamically suppress certain symmetry generators, resulting in a reduced realization of the full $\mathfrak{osp}(1|2)_k$ structure. This kind of toy model construction, where a specific field configuration is used to analyze residual symmetry algebras, is also employed in the context of AdS$_2$ boundary conditions ~\cite{Grumiller:2017qao}.

The dilaton field plays the role of a stabilizer for the $\mathfrak{osp}(1|2)$ gauge connection, interacting with it through boundary fluctuations and thereby obstructing the complete realization of the full infinite--dimensional $\mathcal{N}=1$ $\mathfrak{osp}(1|2)$ symmetry. This symmetry reduction originates from the fluctuating behavior of the dilaton at the boundary. Table~$1$ summarizes the superconformal charges and their corresponding dilaton spin values. Consequently, the residual symmetries encoded in the fields $a_t$ and $x$ manifest as diffeomorphisms on the boundary circle, generated by the vector fields $\epsilon$ and $\eta$.

We therefore identify the ASA as the classical extended $\mathcal{N}=1$ superconformal algebra, rather than its quantum counterpart. This classical algebra arises naturally in our model and is consistent with the symmetry--breaking mechanism induced by the dilaton. Up to this point, our discussion has remained within the classical regime of the extended $\mathcal{N}=1$ superconformal symmetry. The quantum version of the theory, however, has already been presented in Eqs.~\eqref{eq:Q1}--\eqref{eq:Q2}. It is natural to ask whether the quantum theory can also be formulated in terms of a classical extended superconformal symmetry, or whether it necessitates a distinct quantum realization. While this remains a nontrivial question, preliminary evidence suggests that the quantum structure may be realized consistently, preserving the essential features of the classical algebra.

It is important to emphasize that the conformal boundary conditions considered in this section are not merely a technical simplification, but rather encode the infrared dynamics of the boundary theory. While the affine boundary data support a richer set of boundary modes—including commuting currents and higher--spin extensions—the conformal reduction effectively isolates the pseudo--Goldstone mode associated with the breaking of reparametrization symmetry. This reduction aligns with the low--energy behavior of dual SYK--like systems and captures the dominant physical degree of freedom via the Schwarzian effective action. Therefore, the transition from affine to conformal boundary conditions should be viewed as a flow from a general UV--complete description to an IR--dominated effective theory.
\newpage
\subsection{Scope and physical context}\label{scope}

The present work develops the $\mathcal{N}=1$ supersymmetric extension of our earlier analysis of $\mathfrak{sl}(3,\mathbb{R})$ JT gravity~\cite{Ozer:2025bpb}. As in that framework, the asymptotic symmetry structure is derived directly from bulk gauge fields and residual gauge transformations, providing a boundary-independent formulation of the ASA.
In the supersymmetric setting governed by the $\mathfrak{osp}(1|2)$ superalgebra, we maintain this bulk-centered perspective. The ASA is obtained intrinsically from the gauge-theoretic structure, without introducing an auxiliary effective boundary action. The emphasis is therefore placed on the structural origin of asymptotic symmetries within the bulk BF formulation.
The phrase ``beyond the Schwarzian regime'' refers to this methodological standpoint. It signals a shift from boundary action–based reductions toward a direct analysis of gauge symmetries and their asymptotic realization. This perspective is consistent with the strategy established in the non-supersymmetric $\mathfrak{sl}(3,\mathbb{R})$ setting and extended here to the $\mathfrak{osp}(1|2)$ supergravity framework.
In this sense, while our construction remains algebraic in scope, the extended bulk symmetry breaking mechanism within the $\mathfrak{osp}(1|2)$ framework may serve as a foundation for future developments involving boundary dynamics, supersymmetric observables, or holographic models that transcend the traditional Schwarzian description. We believe that this perspective may complement ongoing efforts to understand JT supergravity beyond its standard boundary effective action paradigm.

From a physical standpoint, the dilaton-induced symmetry-selection mechanism analyzed in this work influences the spectrum of boundary observables in putative holographic dual theories. In the affine regime, the full 
$\mathfrak{osp}(1|2)_k$ current structure remains intact, while specific dilaton backgrounds dynamically select a residual subset of symmetry generators together with abelian commuting modes. This background-dependent selection modifies the spectrum of conserved charges and constrains the admissible class of boundary excitations.
By contrast, the Drinfeld–Sokolov reduction discussed in Section~\ref{sec:bhreduction2} isolates the superconformal sector governing the infrared dynamics. In that regime, the pseudo-Goldstone mode associated with super-reparametrization symmetry becomes the dominant boundary degree of freedom, providing the standard low-energy description familiar from super-Schwarzian analyses.
Although an explicit construction of the dual theory lies beyond the scope of this paper, the algebraic structures derived here provide structural constraints for supersymmetric SYK-like models and other one-dimensional superconformal systems. Further exploration of these implications remains an important direction for future work.
\medskip

\noindent
We conclude this section by briefly addressing the quantum aspects associated with the classical analysis presented above. While the central extension of the superalgebra follows naturally from the classical coadjoint 
orbit structure, its precise operator realization at the quantum level requires a consistent treatment of ordering prescriptions. A systematic quantum formulation—such as path integral quantization or BRST analysis—lies beyond the present scope. Nevertheless, the classical results derived here establish a well-defined structural starting point for future investigations, particularly in the context of supersymmetric generalizations of the Schwarzian theory.

\medskip
\noindent
Future work can extend the present analysis to quantum realizations of the symmetry-selection mechanisms identified here, particularly in the context of super-Schwarzian dynamics and SYK-like boundary theories. The toy model introduced in Section~\ref{sec:bhreduction2} offers a controlled setting for probing correlation functions and operator spectra in prospective dual descriptions. More broadly, the structural framework developed in this work provides a basis for further investigations of two-dimensional supergravity and its holographic applications.

\paragraph{Quantum Extensions and Ordering Effects.}

The classical analysis developed in the previous sections admits a natural quantum extension, but this step is accompanied by several structural subtleties. In our framework, central extensions arise geometrically from the coadjoint orbit construction. More precisely, the Kirillov--Kostant symplectic form defined on the orbit determines the classical central term of the $\mathcal{N}=1$ superconformal algebra in a coordinate-independent manner~\cite{Grumiller:2017qao,Ikeda:1993fh}. In this sense, the central charge is not introduced \emph{ad hoc}, but emerges from the underlying symplectic geometry of the reduced phase space.

Upon quantization, however, the transition from Poisson brackets to commutators introduces operator ordering ambiguities. These effects become particularly delicate in superalgebraic systems such as $\mathfrak{osp}(1|2)$, where fermionic generators and second-class constraints coexist. While the classical brackets close consistently, the quantum realization of the symmetry generators requires a careful choice of normal ordering prescription to preserve associativity, supersymmetry, and the graded Jacobi identities.

Several complementary quantization schemes may be employed to address these issues. Coadjoint orbit quantization provides a geometrically transparent construction of the Hilbert space by promoting the orbit symplectic form to a quantum commutator structure~\cite{Grumiller:2017qao,Ikeda:1993fh}. In contrast, BRST quantization systematically incorporates gauge constraints and residual symmetries into a nilpotent cohomological framework~\cite{Ikeda:1993fh,Grumiller:2013swa}, which is particularly well-suited to BF-type supergravity models. Finally, the path integral formulation captures quantum fluctuations around classical backgrounds and makes contact with the super-Schwarzian boundary theory and its partition function~\cite{Engelsoy:2016xyb,Cardenas:2018krd}.

Although a full quantum analysis lies beyond the scope of the present work, the algebraic symmetry breaking mechanism and the Brown--Henneaux reduction discussed in Section~\ref{sec:bhreduction2} provide a controlled starting point for investigating quantum deformations of the boundary algebra. In particular, the affine-to-superconformal reduction suggests a natural framework for studying quantum corrections to the extended $\mathcal{N}=1$ symmetry and its dual one-dimensional superconformal dynamics.
\section{Conclusions and Discussion}\label{sec:conclusion}

Two--dimensional gravity models, most notably JT gravity, provide a remarkably rich arena in which geometry, boundary dynamics, and holography intertwine. In the $\mathfrak{sl}(2,\mathbb{R})$ BF formulation, the theory fixes the asymptotic structure of AdS$_2$ and promotes boundary diffeomorphisms to an infinite--dimensional Virasoro symmetry algebra. Our analysis shows that the realization of this symmetry structure is structurally sensitive to the dynamical role of the dilaton. Once the dilaton develops nontrivial time dependence at the boundary, the set of admissible residual gauge transformations is dynamically restricted: the expected $\St\Lt(2,\mathbb{R})$ subgroup embedded in the Virasoro algebra is no longer generically preserved as a stabilizer of the background configuration. This symmetry selection originates directly from boundary dilaton fluctuations and reflects the fact that the reduced phase space is controlled not only by the metric sector but also by dilaton degrees of freedom. Consequently, the dilaton is elevated from a geometric marker of curvature to a central dynamical variable that governs the boundary realization of asymptotic symmetries.

This foundational mechanism is further developed in the present work within an $\mathfrak{osp}(1|2)$ supergravity extension, where symmetry breaking and symmetry extension are realized simultaneously in a unified gauge--theoretic framework. In this supersymmetric setting, the dilaton supermultiplet not only constrains the admissible residual symmetries but also enriches the boundary algebraic structure, thereby linking affine and superconformal regimes within a single dynamical construction.

As a natural supersymmetric extension of classical $\mathfrak{sl}(2,\mathbb{R})$ JT gravity, 
the $\mathfrak{osp}(1|2)$ JT supergravity framework introduces a dynamically 
enlarged phase space through the inclusion of fermionic generators and the 
dilaton supermultiplet. This extension does not simply amount to a formal 
augmentation of the algebra; rather, it reorganizes the gauge sector and the 
boundary realization of symmetries while preserving the underlying affine 
structure. In particular, the supersymmetric formulation enriches the space 
of admissible residual symmetries and establishes a coherent link between 
affine and superconformal regimes within a unified gauge--theoretic description.

Notably, the $\mathfrak{osp}(1|2)$ gauge formulation reveals a background-dependent 
mechanism of symmetry selection, in which the realization of residual symmetries 
is dynamically constrained by the dilaton supermultiplet. Importantly, this does 
not deform the underlying affine structure, but rather restricts the subset of 
admissible generators compatible with a given boundary configuration. As a 
result, the boundary dynamics acquire additional structure relative to the purely 
bosonic case, while preserving the consistency of the gauge algebra. 
This extended framework provides a controlled setting for exploring generalized 
holographic correspondences in which supersymmetric symmetry realizations influence 
thermodynamic observables such as energy and entropy. In this sense, the theory 
offers a coherent arena for studying the interplay between supersymmetry and 
low-dimensional holography within a unified gauge-theoretic description.

In contrast to metric-based or purely coadjoint orbit formulations of 
$\mathfrak{osp}(1|2)$ dilaton supergravity, the BF-theoretic framework adopted 
here offers a unified gauge-theoretic derivation of the $\mathcal{N}=1$ 
superconformal algebra and its residual symmetry structure. Rather than 
postulating the boundary algebra \emph{a priori}, it emerges directly from the 
analysis of admissible gauge transformations and boundary conditions.
Within this formulation, the dynamical role of the dilaton supermultiplet leads 
to symmetry-selection effects that extend the structure encountered in the 
purely bosonic $\mathfrak{sl}(2,\mathbb{R})$ case. These effects arise naturally 
from the gauge-theoretic treatment and clarify how supersymmetric extensions 
modify the realization of boundary symmetries without deforming the underlying 
affine algebra.

The primary structural distinction between $\mathfrak{sl}(2,\mathbb{R})$ and 
$\mathfrak{osp}(1|2)$ JT gravity lies not in a mere algebraic enlargement, 
but in the dynamical role of the dilaton supermultiplet and its influence 
on the realization of asymptotic symmetries. While the $\mathfrak{sl}(2,\mathbb{R})$ 
formulation captures the isometric structure of $\mathrm{AdS}_2$ geometry 
and its Virasoro enhancement, the $\mathfrak{osp}(1|2)$ framework reorganizes 
the reduced phase space by incorporating fermionic generators and 
background-dependent symmetry selection mechanisms.
This supersymmetric gauge-theoretic setting provides a controlled arena 
for analyzing richer boundary symmetry realizations and their holographic 
implications. In particular, it offers a natural starting point for 
systematic generalizations to higher superalgebras within the BF framework, 
in close analogy with Chern--Simons constructions of three-dimensional gravity.

The results of this work show that the asymptotic symmetry structure is determined 
not only by the gauge connection but also by the dynamical boundary configuration 
of the dilaton field. The dilaton is not an external matter sector; rather, it is 
an intrinsic component of the BF formulation, valued in the adjoint representation 
of the gauge algebra.
Under suitable boundary conditions, time--dependent dilaton profiles dynamically 
select admissible residual symmetry realizations and generate additional commuting 
boundary modes. Importantly, this effect does not arise from introducing an 
external $U(1)$ sector, but from internal gauge degrees of freedom that become 
dynamically relevant near the boundary. The resulting commuting currents should 
therefore be understood as emerging from the intrinsic phase space structure, 
modifying the realization of the ASA without deforming 
its underlying affine structure. 
In this sense, the boundary behavior of the dilaton plays a decisive role in 
shaping the effective symmetry realization, highlighting how internal gauge 
dynamics govern the asymptotic structure of two--dimensional gravitational systems.

In conclusion, the dynamical role of the dilaton field plays a central part 
in shaping the boundary realization of asymptotic symmetries in both 
$\mathfrak{sl}(2,\mathbb{R})$ and $\mathfrak{osp}(1|2)$ JT gravity frameworks. Rather 
than deforming the underlying affine structures, the dilaton modifies the 
set of admissible residual symmetry realizations through background-dependent 
selection mechanisms. 
The supersymmetric extension studied here demonstrates how a unified 
gauge-theoretic formulation naturally accommodates both affine and 
superconformal regimes within a single coherent phase space description. 
This perspective clarifies the structural role of internal gauge dynamics 
in determining boundary observables and their symmetry properties. 
More broadly, such gauge-based constructions provide a systematic framework 
for exploring refined holographic correspondences, including connections to 
super-Schwarzian dynamics and SYK-like models. Extensions to higher-rank 
(super)algebras within the BF formalism offer a natural direction for 
future research in low-dimensional holography.

As discussed in the Introduction, recent studies have explored connections 
between generalized JT gravity, deformed SYK models, and extended symmetry 
structures in low-dimensional holography. In particular, Ref.~\cite{Momeni:2024ygv} 
analyzes deformed SYK-type models within the broader framework of generalized 
dilaton gravity. 
Within this context, the $\mathfrak{osp}(1|2)$ construction developed here 
provides a concrete gauge-theoretic realization of $\mathcal{N}=1$ 
superconformal symmetry together with a controlled mechanism of 
background-dependent symmetry selection. Complementary approaches based on 
coadjoint orbit techniques, quantization of extended superconformal algebras, 
and generalized boundary dynamics have been developed in 
\cite{Momeni:2020tyt,Momeni:2020zkx,Momeni:2021jhk}.
While sharing similar motivations, the present BF formulation differs in that 
the asymptotic symmetry structure is derived directly from admissible gauge 
transformations and boundary conditions, rather than postulated at the level 
of an effective boundary action. This unified gauge perspective clarifies how 
supersymmetric extensions and their associated boundary realizations can be 
consistently embedded within two-dimensional holographic frameworks.

In conclusion, JT gravity formulated within the two--dimensional BF framework 
provides a transparent gauge--theoretic description of the geometric origin 
of asymptotic symmetries and their boundary realizations. The bulk gauge 
symmetries, together with the algebraic structure of the target space, 
determine the admissible boundary transformations and the associated 
ASAs.
Within this setting, the Schwarzian action arises as an effective description 
of a reduced sector of these symmetries, obtained after imposing suitable 
boundary conditions. While it captures the dynamics of the physically relevant 
reparametrization modes, it does not exhaust the full gauge structure present 
in the BF formulation. 
The BF perspective therefore clarifies how JT gravity encodes boundary 
symmetry realization in a systematic and internally consistent manner, 
providing a coherent framework for analyzing their holographic interpretation.

An open question concerns whether the $\mathfrak{osp}(1|2)$ JT gravity 
framework developed here admits a well-defined quantum mechanical dual, 
potentially in the spirit of supersymmetric SYK-like models with extended 
boundary symmetry structures. 
In our earlier $\mathfrak{sl}(3,\mathbb{R})$ analysis~\cite{Ozer:2025bpb}, 
we showed that boundary dynamics can extend beyond the minimal Schwarzian 
sector. In a similar spirit, the present study does not attempt to reproduce 
the standard super-Schwarzian construction~\cite{Cardenas:2018krd}, but 
rather provides a gauge-theoretic setting in which alternative boundary 
realizations may consistently arise.
Within this perspective, generalizations to extended 
$\mathfrak{osp}(\mathcal{N}|2)$ BF theories with $\mathcal{N}>1$ 
offer a natural direction for exploring richer asymptotic symmetry 
realizations in future work.

In addition, we have clarified how the transition from the affine to the 
superconformal regime may be understood as an infrared truncation, 
consistent with the dominance of pseudo--Goldstone boundary modes 
associated with super-reparametrization symmetry. This interpretation 
places the Schwarzian sector within a broader gauge-theoretic framework 
rather than treating it as a fundamental starting point.
We have also outlined how coadjoint orbit, BRST, and path integral 
quantization approaches provide complementary perspectives for 
investigating the quantum realization of the boundary symmetry structure. 
Together, these observations indicate several coherent directions for 
developing a more systematic understanding of the quantum dual 
description.

Although our construction does not introduce an explicit boundary action, 
the residual $\mathfrak{osp}(1|2)$ symmetry structure derived from the bulk 
gauge formulation naturally invites comparison with supersymmetric 
Schwarzian-type quantum mechanical models. The emphasis of the present work, 
however, is on the intrinsic gauge-theoretic realization of boundary symmetries 
rather than on constructing a specific holographic dual.
The algebraic framework uncovered here may nevertheless provide structural 
constraints for prospective boundary effective descriptions consistent with 
the extended supersymmetric ASA. Determining whether 
such realizations reproduce, generalize, or depart from the standard 
super-Schwarzian sector remains an open and well-posed direction for 
future investigation.

The main contribution of this work is the systematic derivation of the 
asymptotic symmetry structure of two--dimensional $\mathcal{N}=1$ dilaton 
supergravity directly from the bulk BF formulation and its residual gauge 
transformations, without postulating an effective boundary action such as 
the super--Schwarzian.
Within this framework, the dilaton supermultiplet does not act as a source 
of algebraic deformation, but rather as a dynamical selector that determines 
the admissible residual symmetry realizations in the asymptotic phase space. 
The resulting symmetry selection should therefore be understood as an intrinsic 
consequence of the bulk gauge structure, rather than as a restriction imposed 
\emph{a posteriori} through holonomy or regularity conditions.
In this way, our analysis complements existing treatments of boundary dynamics 
while providing a structurally refined gauge-theoretic perspective on the origin 
and realization of asymptotic symmetries in supersymmetric dilaton gravity.

\section{Acknowledgments}
\label{sec:ack}
\textit{
Authors are supported by The Scientific and Technological Research Council of Türkiye (TÜBİTAK) through the ARDEB 1001 project with Grant number 123F255. 
}
\appendix
\section{Notation Table}
\label{sec:appa}

\begin{table}[h!]
\centering
\begin{tabular}{|c|l|}
\hline
\textbf{Symbol} & \textbf{Description} \\
\hline
$\Lt_i, \Gt_i$ &  $\mathfrak{osp}(1|2)$ superalgebra basis \\
\hline
$\mathcal{L}^i,\mathcal{G}^i$, $\mathcal{X}^i, \mathcal{Y}^i$ & affine boundary charges and dilaton components \\
\hline
$\epsilon^i, \eta^i$ & affine boundary variation parameters \\
\hline
$\mathcal{L},\mathcal{G}$, $\mathcal{X}, \mathcal{Y}$ & superconformal boundary charges and dilaton components \\
\hline
$\epsilon, \eta$ & superconformal boundary variation parameters \\
\hline
\end{tabular}
\caption{Notation Table}
\end{table}

\section{Summary of Boundary Theories}
\label{sec:appB}

\begin{table}[h!]
\centering
\begin{tabular}{|c|l|c|}
\hline
\textbf{Dilaton gravity theory} & \textbf{Boundary fields} & \textbf{ASA} \\
\hline
\(\mathfrak{sl}(2,\mathbb{R})\)-BF affine & $\mathcal{L}^i, \mathcal{X}^i$ & Extended affine $\mathfrak{sl}(2,\mathbb{R})_k$ \\
\hline
\(\mathfrak{sl}(2,\mathbb{R})\)-BF conformal & $\mathcal{L}, \mathcal{X}$ & Extended $\mathcal{W}_2$ \\
\hline
$\mathfrak{osp}(1|2)$-BF affine & $\mathcal{L}^i, \mathcal{G}^i, \mathcal{X}^i, \mathcal{Y}^i$ & Extended affine $\mathfrak{osp}(1|2)_k$\\
\hline
$\mathfrak{osp}(1|2)$-BF superconformal & $\mathcal{L}, \mathcal{G}, \mathcal{X}, \mathcal{Y}$ & Extended $\mathcal{N}=1$ superconformal  \\
\hline
\end{tabular}
\label{tab:boundary-summary}
\caption{
\shortstack{
Boundary dynamics and asymptotic symmetries.
}
}
\end{table}


\end{document}